\documentclass[aps,prd,floatfix]{revtex4}
\usepackage{epsf}

\newcommand{\bee}{\begin{equation}}
\newcommand{\ee}{\end{equation}}
\newcommand{\beea}{\begin{eqnarray}}
\newcommand{\eea}{\end{eqnarray}}

\begin{document}
\title{Kaon B Parameter in Quenched QCD}

\author{Thomas DeGrand}
\affiliation{
Department of Physics, University of Colorado,
        Boulder, CO 80309 USA}

\author{MILC Collaboration}
\noaffiliation

\date{\today}

\begin{abstract}
I calculate the kaon B-parameter $B_K$, defined via
${8\over 3} (m_K f_K)^2 B_K =\langle \bar K| \bar s \gamma_\mu(1-\gamma_5) d
\bar s \gamma_\mu(1-\gamma_5) d | K \rangle $, with a lattice simulation
in quenched approximation.  The lattice simulation uses an action possessing exact
 lattice chiral symmetry, an overlap action.
Computations are performed at two lattice spacings, about 0.13 and 0.09 fm
(parameterized by Wilson gauge action
couplings $\beta=5.9$ and 6.1) with nearly the same physical
volumes and quark masses.
I describe  particular potential difficulties which arise due to the use of
such a lattice action in finite volume. My results are consistent
with other recent lattice determinations using domain-wall fermions.

\end{abstract}
\maketitle

\section{Introduction}
The kaon B-parameter $B_K$,  defined as
${8\over 3} (m_K f_K)^2 B_K =\langle \bar K| \bar s \gamma_\mu(1-\gamma_5) d
\bar s \gamma_\mu(1-\gamma_5) d | K \rangle $, is an important
 ingredient in the testing
of the unitarity of the Cabibbo-Kobayashi-Maskawa matrix \cite{Hocker:2001xe}.
It has been a target of lattice calculations since the earliest
 days of numerical simulations of QCD.
Lattice calculations of $B_K$ require actions with good chiral properties,
since the matrix element of the four-fermion operator scales like the
 square of the 
pseudoscalar meson mass as that mass vanishes. If the lattice action does not
respect chiral symmetry, the desired operator will mix with operators
 of opposite chirality.
The matrix elements of these operators do not vanish at vanishing
 quark mass, and therefore  overwhelm the signal.

 There has been a continuous  cycle of lattice calculations using
 fermions with ever better chiral properties.
This calculation is yet another incremental upgrade, to  the use
of a lattice action with exact $SU(N_f)\otimes SU(N_f)$ chiral symmetry,
an overlap \cite{ref:neuberfer}
action. These actions have operator mixing identical to that
of continuum-regulated QCD \cite{Capitani:2000da}.

The first lattice calculations of $B_K$ were done with Wilson-type 
actions. Techniques
for handling operator mixing have improved over the years, (for  recent results,
see \cite{Gupta:1996yt}-\cite{Aoki:1999gw})
 but this approach remains (in the author's opinion) arduous.

Staggered fermions (\cite{Kilcup:1997ye}--\cite{Aoki:1997nr})
have enough chiral symmetry at nonzero lattice spacing, that
operator mixing is not a problem. One can obtain extremely precise
 values for lattice-regulated
 $B_K$ at any fixed lattice spacing. However, to date, all calculations
 of $B_K$ done
 with staggered fermions use ``unimproved'' (thin link,
 nearest-neighbor-only interactions),
and scaling violations are seen to be large. For example,
 the JLQCD collaboration \cite{Aoki:1997nr}
saw a thirty per cent variation in $B_K$ over their range
 of lattice spacings.

Domain wall fermions  pin chiral fermions to a four
 dimensional brane in a five dimensional world; 
chiral symmetry is exact as the length of the fifth 
 dimension becomes infinite. For real-world
simulations, the fifth dimension is finite and chiral
 symmetry remains approximate, though
much improved in practice compared to Wilson-type fermions. Two groups 
\cite{AliKhan:2001wr,Blum:2001xb} have presented results
 for $B_K$ with domain wall fermions.
Ref. \cite{AliKhan:2001wr} has data at two lattice
 spacings and sees only small scaling violations.
There is a few standard deviation disagreement
 between the published results of the two groups.

Finally, overlap actions have exact chiral symmetry
 at finite lattice spacing. All operator
mixing is exactly as in the continuum \cite{Capitani:2000da}.  Two
groups \cite{Garron:2003cb,DeGrand:2002xe} have recently
 presented  results for $B_K$
using overlap actions, but the actions and techniques
 are completely different.
The second of these is a preliminary version of the work described here.

The lattice matrix elements must be converted
 to their continuum-regularized values and run to
some fiducial scale. Matching coefficients
 can be computed perturbatively or nonperturbatively.
For most standard discretizations of fermions,
 the matching factors (``Z-factors'')
are quite different from unity. For these actions,
perturbation theory is regarded as untrustworthy, and other
methods must be employed. I, however, use an action in which the gauge
connections  are an average of a set of short range paths,
 specifically HYP-blocked
 links\cite{ref:HYP}.
Ref. \cite{DeGrand:2002va} has computed the
 matching factors for operators relevant for this study, as well as
the scale for evaluating the running coupling, using the
 Lepage-Mackenzie-Hornbostel
\cite{ref:LM,Hornbostel:2000ey} criterion. The Z-factors
 are quite close to unity.
A number of calculations of Z-factors for related actions \cite{Bernard:1999kc}
reveal that this behavior is generic for actions with
 similar kinds of gauge connections.
 In this work I compare perturbative and nonperturbative calculations of  two
 Z-factors, for the local
 axial current and for the lattice-to-
$\overline{MS}$ quark mass matching factor, and 
find reasonable agreement between them.

Having exact chiral symmetry forces one to confront the problem
that this calculation is performed in  quenched approximation.
These results will not
be directly applicable to the real world of QCD with
 dynamical fermions.
I encounter difficulties in two places. First, there is no
reason for the spectrum of quenched QCD to be identical to that
of full QCD. Using different physical observables to extract
a strange quark mass can (and does) lead
to different values of this parameter. At small quark masses,
my observed $B$ parameter varies strongly with quark mass,
and my prediction for $B_K$ is sensitive to my choice of $m_s$.

The second difficulty is more fundamental. In order
to extrapolate  results
to the chiral limit, one must use chiral perturbation theory.
However, the symmetries of quenched QCD are different from the symmetries of full
QCD.
Not only are the leading coefficients expected to be different in quenched
and full QCD, but the logarithmic contributions to any observable $Q$
\bee
Q(m_{PS}) = A(1 + B{ m_{PS}^2 \over f_{PS}^2 }\log m_{PS}^2 )+ \dots
\label{eq:eq001}
\ee
can have different coefficients (different $B$'s in Eq. \ref{eq:eq001}),
or different functional form (in the formula for $m_{PS}^2/m_q$, the
coefficient of $\log(m_{PS}^2)$ is not $m_{PS}^2$ but a constant related to
the quenched topological susceptibility \cite{Sharpe:1992ft,BG}).  All of these differences are
encountered in the analysis of the data. To produce a prediction for
an experimental number using results of a quenched simulation involves
uncontrolled, non-lattice-related phenomenological assumptions.

\begin{figure}[t!hb]
\begin{center}
\epsfxsize=0.6 \hsize
\epsffile{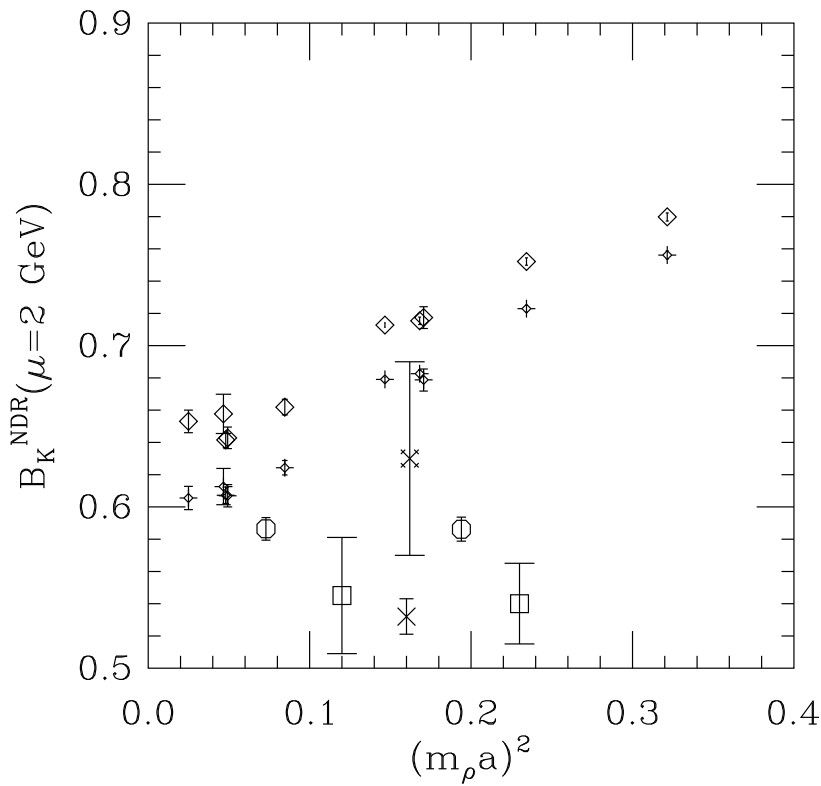}
\end{center}
\caption{
$B_K$ comparisons vs lattice spacing, from a selection of simulations
with reasonably small error bars. Results are labeled
diamonds and fancy diamond{\protect{\cite{Aoki:1997nr}}},
the fancy cross {\protect{\cite{Garron:2003cb}}},
octagons {\protect{\cite{AliKhan:2001wr}}},
the cross {\protect{\cite{Blum:2001xb}}},
and squares (this work).
}
\label{fig:worldbka}
\end{figure}

\begin{figure}[t!hb]
\begin{center}
\epsfxsize=0.6 \hsize
\epsffile{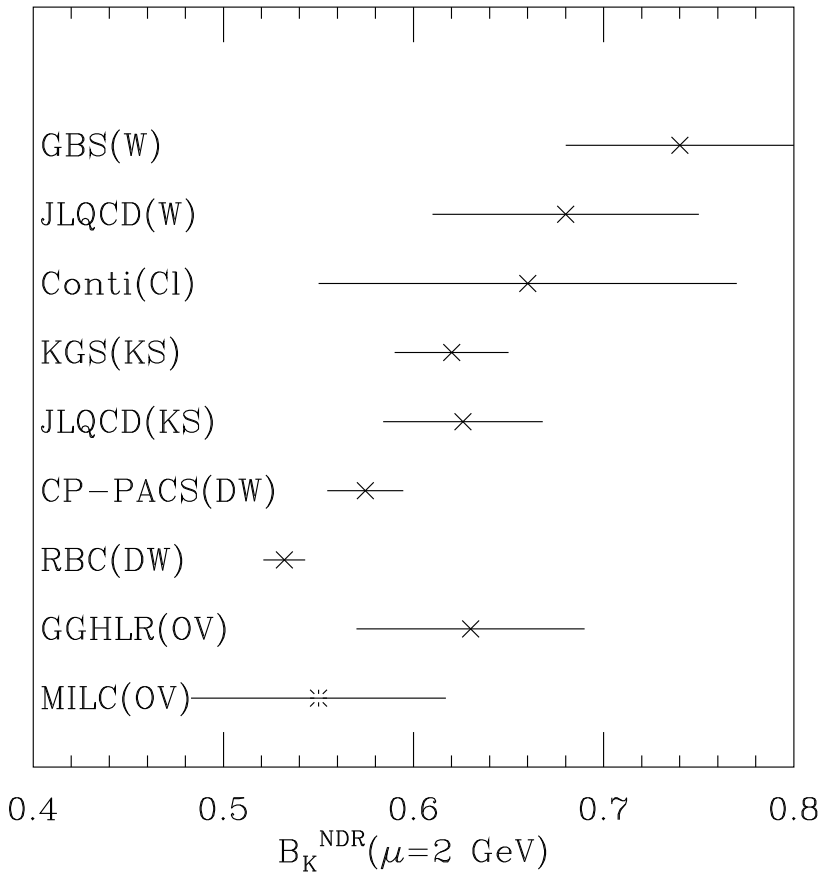}
\end{center}
\caption{
$B_K$ comparisons presented ``as if'' they were taken to the continuum limit.
The label in parentheses characterizes the kind of lattice fermions used:
W for Wilson, Cl for Clover, KS for staggered, DW for domain wall, and
OV for overlap fermions.
References are
GBS \protect{\cite{Gupta:1996yt}},
JLQCD(W) \protect{\cite{Aoki:1999gw}},
Conti \protect{\cite{Conti:1997qk}},
KGS \protect{\cite{Kilcup:1997ye}},
JLQCD(KS) \protect{\cite{Aoki:1997nr}},
CP-PACS \protect{\cite{AliKhan:2001wr}},
RBC \protect{\cite{Blum:2001xb}},
GGHLR \protect{\cite{Garron:2003cb}},
and MILC, this work.
The points of Refs.
\protect{\cite{Aoki:1999gw}},
\protect{\cite{Kilcup:1997ye}},
\protect{\cite{Aoki:1997nr}},
\protect{\cite{AliKhan:2001wr}},
and this work are the results of a  a continuum extrapolation; all the rest are
 simulations at one lattice spacing.
}
\label{fig:worldbk}
\end{figure}

I conclude this introduction by presenting graphs which
 illustrate my results.
Fig. \ref{fig:worldbka} shows results for $B_K$ at various lattice spacings,
for a selection of
simulations which have reasonable statistics and small error bars.
Fig. \ref{fig:worldbk} presents results which are either extrapolated
to the continuum limit, or presented by their authors as having
 small lattice spacing artifacts.

In Section 2 I describe the action, simulation parameters, and data sets.
Sec. 3 is devoted to a discussion of zero mode effects and my attempts
 to deal with them.
Results relevant to $B_K$ are presented in Sec. 4.
In Sec. 5 I discuss the chiral limit of $B_K$ and of
operators $O_7^{3/2}$ and $O_8^{3/2}$, relevant for part
of $\epsilon'/\epsilon$. My brief conclusions are given in Sec. 6.
An Appendix compares perturbative and nonperturbative matching factors.
\section{Simulation techniques}

\subsection{Data sets}
The data set used in this study  is generated in the quenched approximation
 using the Wilson gauge action at  couplings
 $\beta=5.9$ (on a $12^3 \times 36$ site lattice), where I have
an 80 lattice data set,
and $\beta=6.1$ (on a $16^3 \times 48$ site lattice)
with 60  lattices.
 The nominal lattice spacings are $a=0.13$ fm and 0.09 fm
from the measured rho mass.
Propagators for ten ($\beta=5.9$) or nine ($\beta=6.1$)
 quark masses are constructed
corresponding to pseudoscalar-to-vector meson mass ratios
of $m_{PS}/m_V$ ranging from 0.4 to 0.85.
The fermions have periodic boundary conditions in the
spatial directions and anti-periodic temporal boundary conditions.
I gauge fix to Coulomb gauge and take our
 sources to be Gaussians of size $x_0/a=3$, 4.125 at $\beta=5.9$, 6.1
(where the quark source is $\Phi = \exp(-x^2/x_0^2)$).

\subsection{Lattice action and simulation methodology}

The massless overlap Dirac operator is
\bee D(0) = x_0(1+ {z \over{\sqrt{z^\dagger z}}} )
\label{eq:gw}
\ee
where $z = d(-x_0)/x_0 =(d-x_0)/x_0$ and $d(m)=d+m$ is a massive ``kernel''
 Dirac operator for mass $m$.
 The massive overlap Dirac operator is
conventionally defined to be
\bee
D(m_q) = ({1-{m_q \over{2x_0}}})D(0) + m_q
\ee
and it is also conventional to define the propagator so that the chiral
modes at $\lambda=2x_0$ are projected out,
\bee
\hat D^{-1}(m_q) = {1 \over {1-m_q/(2x_0)}}(D^{-1}(m_q) - {1\over {2x_0}}) .
\label{SUBPROP}
\ee
This also converts local currents into order $a^2$ improved operators \cite{Capitani:1999uz}.

The overlap action used in these studies\cite{ref:TOM_OVER}
 is built from a kernel action with nearest and
next-nearest neighbor couplings,
and  HYP-blocked links\cite{ref:HYP}.
HYP links fatten the gauge links
without extending gauge-field-fermion couplings beyond a single hypercube.
This improves the kernel's chiral properties without compromising locality.

The ``step function'' ($\epsilon(z)=z/\sqrt{z^\dagger z}$) is evaluated using
 the fourteenth-order Remes
algorithm of Ref. \cite{ref:FSU} (after removing the lowest
20 eigenmodes of $z^\dagger z$). This involves an inner
 (multimass \cite{ref:multimass})
conjugate gradient inversion step. It is convenient to monitor the 
norm of the step function
$|\epsilon(z)\psi|^2/|\psi|^2$ and adjust the conjugate gradient
 residue to produce a desired
accuracy  (typically $10^{-5}$ in $\epsilon(z)\psi$). Doing so,
I need about 16-18 inner conjugate gradient steps at $\beta=5.9$
and 10-12 steps at $\beta=6.1$.

An important ingredient of this overlap program has been to precondition the
quark propagator by projecting  low eigenfunctions of the Dirac operator out of
the source and including them exactly. This can in principle
 eliminate critical slowing
down from the iterative calculation of the inversion of the Dirac operator.
Of course, there is a cost: one must find the eigenmodes. My
 impression from the
literature, plus my own experience, is that for the overlap with
 Wilson action kernel,
this cost is prohibitive. However, my kernel action is designed to resemble the
 exact overlap well enough
that its eigenvectors are good ``seeds'' for a calculation of
eigenvectors of the exact action, and it is kept simple enough that
finding its own eigenvectors is inexpensive. 

\begin{figure}[thb]
\begin{center}
\epsfxsize=0.6 \hsize
\epsffile{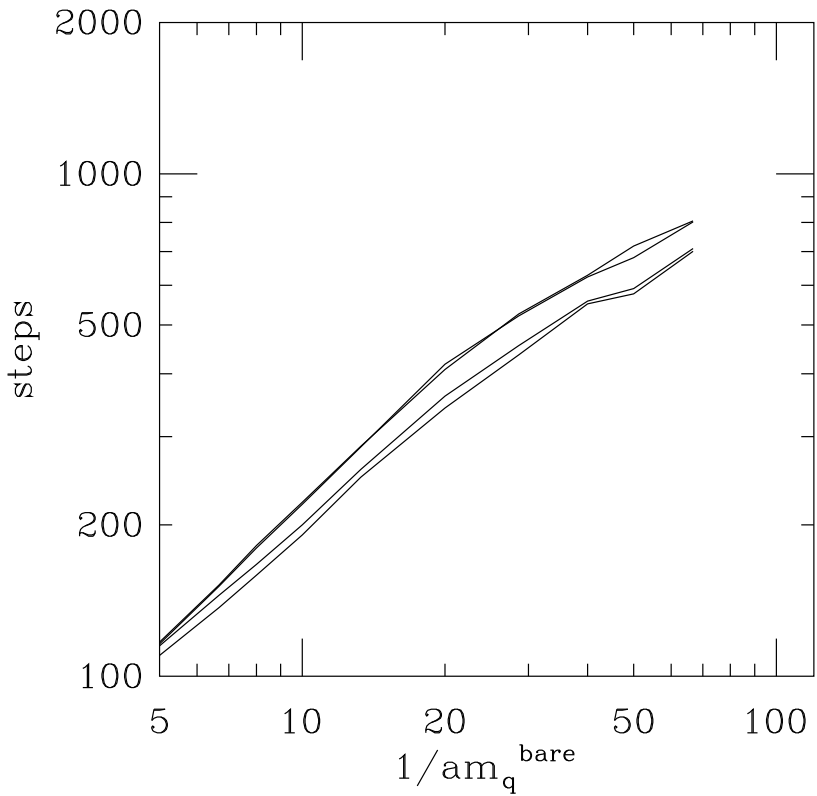}
\end{center}
\caption{
Number of conjugate gradient steps in the calculation of quark
propagators, as a function of quark mass. The four curves are two
 different sources
on each of two different lattices, and we are converging to a
 fractional accuracy of the
squared residue of $10^{-14}$ for this test case.
}
\label{fig:cgsteps}
\end{figure}

As a rough figure of merit, consider the $12^3\times 36$, $\beta=5.9$ data set.
 Computing the lowest twenty eigenmodes of the squared massless 
overlap Dirac operator takes about 8 time units, while the complete set of
 quark propagators
from the lightest mass I studied to the heaviest takes about 16 time
 units times two (for two sets
of propagators) per lattice. Fig. \ref{fig:cgsteps} shows the number of conjugate gradient 
steps needed to compute the quark propagator to some fixed accuracy, as a function of bare quark mass.
If one would naively extrapolate the heavy quark results (for which low eigenmodes
make only a small contribution) to lower masses, one would see a  forty per cent reduction
 in the number of inverter steps needed at the
smallest quark mass, vs an additional cost of 8/32 = 25 per cent per propagator set
for the construction of eigenmodes.

%

\subsection{Correlation functions}
The ``generic'' four fermion operator one must consider is
\bee
O = ( \bar q^{(1)}_\alpha \Gamma_1  q^{(2)}_\beta) \otimes
( \bar q^{(3)}_\gamma \Gamma_2 \hat  q^{(4)}_\delta)
\ee
(the superscript labels flavor; the subscript, color).
Special cases are (a) 
 $O=O_1$: $\Gamma_1=\Gamma_2 = \gamma_\mu(1-\gamma_5)$,
 $\alpha=\delta$, $\beta=\gamma$; 
(b) $O=O_2$:  $\Gamma_1=\Gamma_2 = \gamma_\mu(1-\gamma_5)$,
 $\alpha=\beta$, $\gamma=\delta$;
and  (c) the isospin 3/2 operators for electroweak penguins,
here written with the normalization conventions used
 in Refs. \cite{Blum:2001xb}
and \cite{Noaki:2001un}
\bee
O_7^{3/2}={1\over 2} \big( (\bar s_\alpha \gamma_\mu(1-\gamma_5) d_\alpha)
[(\bar u_\beta \gamma_\mu(1+\gamma_5) u_\beta)
-(\bar d_\beta \gamma_\mu(1+\gamma_5) d_\beta)]
+ (\bar s_\alpha \gamma_\mu(1-\gamma_5) u_\alpha)
(\bar u_\beta \gamma_\mu(1+\gamma_5) d_\beta) \big)
\label{eq:o7}
\ee
and
\bee
O_8^{3/2}={1 \over 2} \big( (\bar s_\alpha \gamma_\mu(1-\gamma_5) d_\beta)
[(\bar u_\beta \gamma_\mu(1+\gamma_5) u_\alpha)
-(\bar d_\beta \gamma_\mu(1+\gamma_5) d_\alpha)]
+ (\bar s_\alpha \gamma_\mu(1-\gamma_5) u_\beta)
(\bar u_\beta \gamma_\mu(1+\gamma_5) d_\alpha)  \big) .
\label{eq:o8}
\ee
 $B_K$ is proportional to the matrix element of  $O_+=O_1+O_2$.
Because overlap fermions are chiral, one can extract $B_K$ from the matrix
element of the operator between zero momentum states.
All the operators to be studied have only ``figure-eight'' topology matrix elements, where
 each field in the operator
contracts against a field in the source or sink interpolating field.
 There are no penguin graphs (where fields in the operator contract against each
 other) as long as one works in the degenerate-mass limit.

Broadly speaking, the matrix element of a four-fermion operator is computed
by placing interpolating fields for a meson at two widely-separated
 locations on 
the lattice and contracting field variables between these sources
and the operator, to construct an
 un-amputated correlator
containing the operator.
There are two commonly-used strategies for doing this: One possibility is to
build all the quark propagators ``at the operator'' at one location on
 the lattice, and
to join up pairs of propagators to make the mesons. The second method
 is to construct
propagators from two well-separated sources and bring them together at
 the operator.
 The advantages of the first
method are that one needs half as many propagators per lattice, and it
 is possible to
project the whole calculation into a particular momentum eigenstate
 (by appropriately
summing over locations of the meson interpolating fields). The major
disadvantage of this method is that one is only measuring the matrix
element of the operator at one location per propagator construction. With
the second method one can average the location of the operator over all
spatial  and many temporal locations, a considerable gain in statistics.
A disadvantage of the second method is that unless
the source generates a hadronic state which is a momentum eigenstate
the correlator will involve a mixture of the eigenstates. The
dominant contribution to the
correlator is from $\vec p=0$, but there will be a contamination from
 higher momentum states.
I have chosen the second method, and will discuss below how I
dealt with higher-momentum modes.
I placed the two source time slices  $N_t/2 -2$ temporal sites apart;
 with the toroidal geometry
there are two temporal regions where the operator is ``between'' the sources.
I combine all two-point function data sets from the two sources in the fits.

The correlator of two interpolating fields located at $x,t=(0,0) $ and $(0,T)$
with an operator $S(x,t)$ summed over $x$ is then
\bee
C_3(t,T)= \sum_x \langle  \Phi(0,T) S(x,t) \Phi(0,0) \rangle
\ee
Inserting complete sets of relativistically normalized momentum eigenstates,
and assuming $\langle h(\vec p_1)|S(\vec p)|h(\vec p_2)\rangle$ is proportional
 to $\delta(\vec p+\vec p_2-\vec p_1)$,
\bee
C_3(t,T)= \sum_{\vec p} \exp(-E(\vec p)T)\langle 0|\Phi |h(\vec p)\rangle|^2
 {1\over {(2E(\vec p))^2}} {1\over V}
  \langle h(\vec p)|S|h(\vec p)\rangle.
\ee
If the $\vec p=0$ state dominates the sum,
\bee
C_3(t,T)= \exp(-mT)|\langle 0|\Phi |h\rangle|^2 {1\over {(2m)^2}} {1\over V}
  \langle h|S|h\rangle  .
\ee

However, ``dominance in the sum'' is controlled by the quantity
 $\exp((-E(\vec p)-m)T)$.
These sources do not make  momentum eigenstates, and so
the $\vec p=0$ $B_K$ signal is contaminated by a $\vec p \ne 0$ contribution.
The minimum momentum in a finite box of side $L$ is $|p|=2\pi/L$.
Because of the small masses involved, the contamination is not present
at  quark masses below 2-3 times the strange
quark mass. It appears at bigger quark mass, because $E(\vec p)-m$ gets
smaller as the pseudoscalar mass $m$ grows.
Fortunately, there are two inequivalent paths on the torus to disentangle
the two ``signals,'' and one can fit the $B_K$ correlator to a sum
of a $\vec p=0$ term and a $p=2\pi/L$ term.

\subsection{Fitting and Error Analysis}
In a Monte Carlo simulation different quantities measured on the same
 set of lattices are
highly correlated, and it is preferable to analyze the data using
 covariant fits. However,
in order to determine accurately the small eigenvalues of the
 correlation matrix, large
statistical samples are needed. If the statistical sample is too small,
 these eigenvalues can be
 poorly determined, and the fits will become unstable. This can be a
 particular problem in
fits which extract $B_K$, simply because they can involve many
 degrees of freedom (one three-point
function and up to two two-point functions). My solution to
 this problem is to
adopt the strategy of the analysis used in  Ref. \cite{Bernard:2002pc}.
 I compute the
correlation matrix (the covariance matrix, but normalized by the
 standard deviation so that its
diagonal entries are the identity), and find its eigenvalues
 and eigenvectors. I then reconstruct the
matrix, discarding eigenvectors whose eigenvalues are smaller
 than some cutoff. This matrix
is singular, so I reset the values of its diagonal entries to
 the identity again. I use
this processed covariance matrix as an input to the fit.

This is basically the strategy adopted in Ref. \cite{Bernard:2002pc},
 except that here
the cut on kept eigenvectors is chosen to
be a fraction of the largest eigenvector, rather than
a fixed value. I have taken the value of this cutoff to be 0.1,
 and have varied it to insure that  fits are insensitive to it.

(My experience in this work is that all fits to spectroscopic
 quantities, and fits to pairs
of two point function, used, for example, to extract pseudoscalar
 decay constants, produce
statistically identical results regardless of the number of
 eigenvalues discarded.)

All chiral extrapolations of data are done by performing correlated
 fits to extract
necessary physical parameters quark mass by quark mass, then
 performing extrapolations
under a single-elimination jackknife.

I have performed a number of different analyses of the data.
Fits of the three point function with a particular sources
(pseudoscalar or axial vector, labeled PS or A) can be
combined with two-point functions with axial vector
sinks to give the B-parameters directly,
\bee
C_3(t,T)= {8\over 3}{{Z^2}\over{V}} (B  \exp(-mT) + C_1 \exp(-E(p)T))
\ee
($E(p)^2= m^2+(2\pi/L)^2$ is the energy of the lowest nonzero momentum state)
and
\bee
C_2(t,t')= Z (\exp(-m(t-t')-\exp(-m(N_t-(t-t')).
\ee
This is done by  performing a correlated fit with three ($Z$, $B$, $m_{PS}$)
or four (add $C_1$) parameters to $C_3(t,T)$, $C_2(t,0)$, and $C_2(t,T)$.

As a check, one can also simply perform a
 (jackknife-averaged) fit of the ratio
 $C_3(t,T)/[C_2(t,0)C_2(t,T)]$ to a constant.
The range of $t$ is varied with a search for plateaus in the fit value.
This method has been commonly used in many
 previous measurements of $B$ parameters. When I compare this
``ratio'' fit to a full correlated fit to $B_K$,
I always find consistency of the fitted $B$ parameter. Generally,
the error bar assigned to it by the ratio fit is about
 forty percent ot the size of
the uncertainty of the correlated fit. However, I elect NOT to use these
more optimistic estimations of the uncertainty because they
do not attempt to include any of the correlations which are known to be
present in the data.

One can also extract the matrix element of the operator by doing a
correlated fit of $C_3(t,T)$ for a particular source and
a two-point function which uses the same source as a sink. In this case
\bee
C_3(t,T)= {{Z'}\over{(2m)^2V}}
(\langle O \rangle  \exp(-mT) + C'_1 \exp(-E(p)T) )
\label{eq:p32a}
\ee
and
\bee
C_2(t,t')= {{Z'} \over{2m}} (\exp(-m(t-t')+\exp(-m(T-(t-t')).
\label{eq:p32b}
\ee
With these correlators, the B parameter can be
extracted by doing a simultaneous fit to another correlator set, which
gives $f_{PS}$ and $m_{PS}$,
and jackknife averaging. This increases the resulting 
uncertainty in $B_K$, so this method
is not competitive with the three-propagator fits.

At lower quark mass the size of the $\vec p \ne 0$
 contamination falls to zero. Fits which include the $C_1$ coefficient
become unstable, because the fit is completely insensitive to it.
The Hessian matrix develops a near-zero eigenvalue as a result of this
insensitivity. Usually, this near zero mode just means that the $C_1$
coefficient will be poorly determined, but a zero mode in a Hessian matrix
is always  dangerous, and can corrupt the whole fit.
 To determine the minimum quark mass
where $C_1$ is not needed, I performed fits where the inversion
 of the covariance matrix was done using singular value decomposition (clipping out
eigenmodes with eigenmodes smaller than some minimum cut). I varied the ratio
 of smallest to largest eigenvalues
kept in the $4 \times 4$ Hessian matrix. Any ``reasonable'' choice
 (conditioning number cut
$\ge 10^{-5}$) stabilizes the inversion  by  decoupling and
freezing out any variation in $C_1$.

\section{Zero mode effects}
Simulations with lattice fermions possessing an exact chiral symmetry
 done in finite
volume have a ``new'' kind of finite volume artifact: the presence of
 exact zero modes
of the Dirac operator, through which quarks can propagate.
In some cases (the eta prime channel, for example) the zero modes
 contribute physics,
but for the case of $B_K$ they are merely an annoyance which would
 disappear in the infinite 
volume limit.

Things could be much worse.
For example, for
 Wilson-type fermions the analogs of zero modes are configurations
where the Dirac operator has real eigenvalues. Because these
theories are not chiral,  eigenvalues of the massless Dirac
 operator $D$ are not ``protected''-- they
can take on any real value. It can happen that this
 real value  coincides with the
negative of the simulation mass $m$. Then $D+m$ is simply
 non-invertible.  Typically, these eigenmodes occur at small quark mass,
meaning that it is difficult (if not
impossible) to
do simulations there.

Note that zero modes will also be present in simulations with nonzero
 mass dynamical fermions,
although  their effects are likely to be less severe; the dynamical fermions suppress
 zero modes but
(except at zero quark mass) do not eliminate them.

There are basically five observables needed in a $B_K$ calculation.
\begin{itemize}
\item{The pseudoscalar mass}
\item{The vector meson mass (to set the lattice spacing)}
\item{The matrix element of $O_+$}
\item{The matrix element of the axial vector current}
\item{The pseudoscalar decay constant}
\end{itemize}
All can (in principle) be affected by finite volume zero
 modes artifacts at small quark mass.

The most obvious way to check for finite volume effects is to
do simulations at several volumes. I did not do that.
One could use sources which do not couple to zero modes. This is the strategy
pursued by Ref. \cite{Garron:2003cb}: their meson interpolating fields are
 the linear
combination $\gamma_0 \gamma_5 - \gamma_0$. I did not do that either.
The $\gamma_0$ part of the source couples only to the (heavy)
 scalar meson, not to the pseudoscalar meson. Its contribution
to the pseudoscalar meson matrix element of the operator
averages to zero but contributes noise in a finite data set.
Because of the PCAC relation, the $\gamma_0 \gamma_5$ source
 decouples for the pseudoscalar meson in the
chiral limit, so that while one is  decoupling zero modes, we
 are also decoupling from the desired channel.

Instead, I used the fact that zero mode effects are different
 in different channels. I measured the same physical
 observable using different operators (presumably with different zero mode
contributions) and looked for differences.

\begin{figure}[thb]
\begin{center}
\epsfxsize=1.0 \hsize
\epsffile{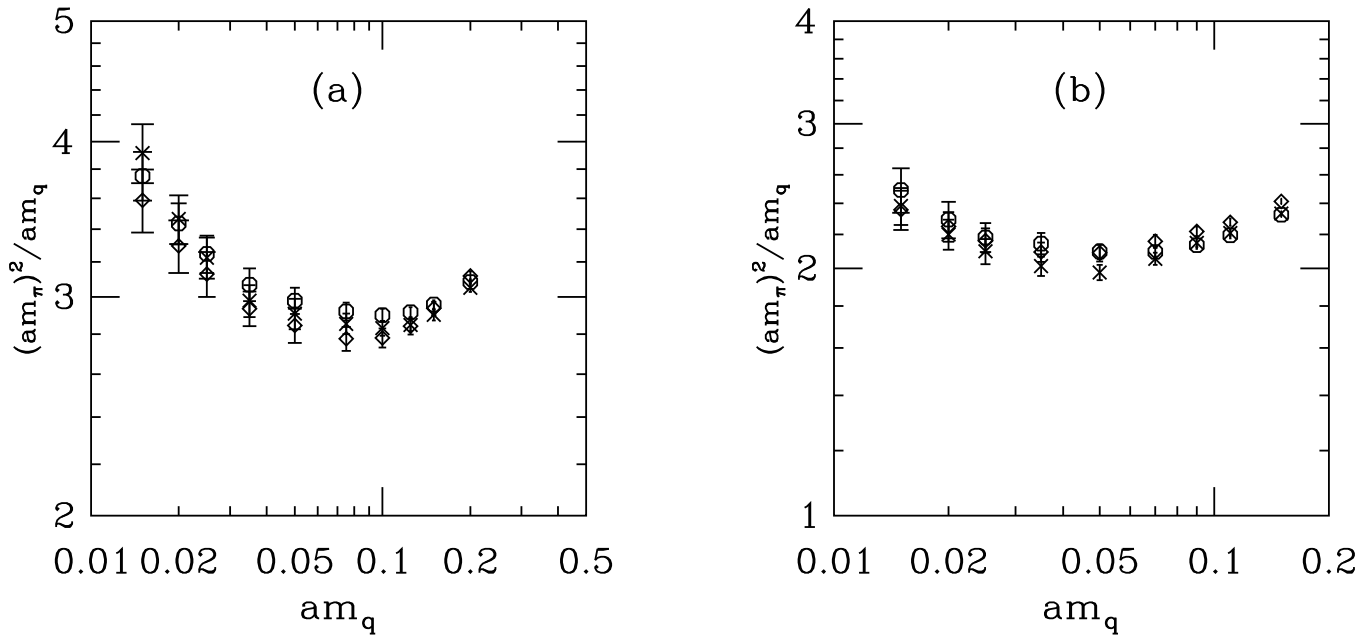}
\end{center}
\caption{ (a) $\beta=5.9$ and (b) $\beta=6.1$:
$(am_{PS})^2/(am_q)$ vs $am_q$ from pseudoscalar (crosses),
 axial vector (octagons)
and pseudoscalar-scalar difference (diamonds) sources.
}
\label{fig:mpi2mqmq}
\end{figure}

 When one fits correlators
to extract pseudoscalar masses, one can perform the fits naively, without
consideration of the zero modes. Then if  zero modes are present
in some channels,
observables such as the pseudoscalar
mass inferred from these channels will show characteristic
differences. As the zero mode correlator reflects some underlying
 structure of the gauge field
responsible for configurations which support zero modes,
 it produces correlations on a scale which is independent of the
quark mass. Thus, a plot of $m_{PS}^2/m_q$ would diverge
at small $m_q$ inversely with $m_q$ (simply because $m_{PS}$ would be independent of $m_q$).
 I searched for this suspicious behavior
in plots of $m_{PS}^2/m_q$ vs $m_q$ (Fig. (\ref{fig:mpi2mqmq}).
Correlators include the pseudoscalar current (which has a contribution
from the propagation of both quarks through zero modes), the
 pseudoscalar-scalar difference (in which all zero mode contributions cancel),
 and the
axial current correlator (which has a ``mixed'' contribution with one quark
propagating through zero modes and the other through nonzero modes:
this contribution in independent of $m_q$ at small $m_q$.)
Within  statistical uncertainties, I do not see any
 difference between these channels.
I anticipate that this would not be the case at smaller quark mass or smaller volumes.
Nevertheless, on theoretical grounds  I elect to  only extract pseudoscalar masses
at small quark masses from the pseudoscalar-scalar difference.

At larger quark mass the pseudoscalar-scalar difference
 degrades as a pseudoscalar source,
 compared to
pseudoscalar or axial currents. The window of timeslices
 in which the fitted mass is stable shrinks.
Presumably what is happening is that the heavier
 pseudoscalar feels the effect of the scalar meson
(with which it would become degenerate in the heavy quark limit).
This effect has been noted by Ref. \cite{Gattringer:2002sb}.

Note (parenthetically) that the data strongly hint at the presence
of an increase in $m_{PS}^2/m_q$ at small $m_q$ which
 is consistent with a quenched chiral
 logarithm. The data set is not optimized to precisely pin down these
effects (the smallest quark mass is still rather large).
In an attempt to look for these terms, I have fit
 $m_\pi^2/m_q$ to two functional forms:
The first one is \cite{Sharpe:1992ft,BG}:
\bee
(am_{PS})^2/(am_q) = A [1 - \delta (\ln ( A m_q/\Lambda) +1)] + B (am_q)
\label{eq:log}
\ee
It contains three free parameters, $A$, $B$, and $\delta$,
 as well as the scale for the chiral
logarithm, $\Lambda$, which I do not let float in the
 fit but instead pin to any of several
fiducial values between 0.8 and 1.2 GeV.

The second functional form  is due to Sharpe \cite{Sharpe:1992ft}:
\bee
(am_{PS})^2/(am_q) = C (am_q)^{(\delta/(1+\delta)}  + D (am_q)
\label{eq:power}
\ee
I elect to perform fits to a mix of pseudoscalar-scalar
 difference correlators at low
quark mass and axial correlators at high mass (rejecting
 the pure pseudoscalars on theoretical grounds).
I saw little shift in the fitted $\delta$ as I varied $\Lambda$, and
the fits give $\delta= 0.17(5)$ at $\beta=5.9$,
0.20(7) at $\beta=6.1$.
The power law fit gives $\delta=0.18(5)$ at $\beta=5.9$
and $0.21(9)$ at $\beta=6.1$.
(All chiral extrapolations from the $\beta=6.1$ data set are noisier than their $\beta=5.9$ 
counterparts because the former set does not extend to as low a value
of the quark mass.)

Recent determinations of $\delta$ from a variety of actions
 show quite wide spread
(for a recent review, see \cite{Wittig:2002ux}).
I am very much aware that my results have uninterestingly
 large error bars compared to the recent high-statistics, low
pseudoscalar mass results of Ref. \cite{Dong:2003im}.
My purpose in quoting them here is to illustrate that I see 
behavior in the data which is consistent with the expectations
of quenched QCD, and inconsistent with the expectations of
full QCD. The choices for the quark mass relevant to the kaon
will occur in the rise near the minimum of the $m_{PS}^2/m_q$ vs $m_q$ plot.

These results are quite a bit greater than ones reported by
 Heller and me \cite{DeGrand:2002gm},
 using an overlap action with
the same fermion kernel, but with gauge connections which
 are much more smoothed than the
ones used in this study. Presumably the smaller value for
 $\delta$ seen there came about
because the additional smoothing suppressed the coupling 
of the fermions to smaller topological
objects.

\begin{figure}[thb]
\begin{center}
\epsfxsize=0.8 \hsize
\epsffile{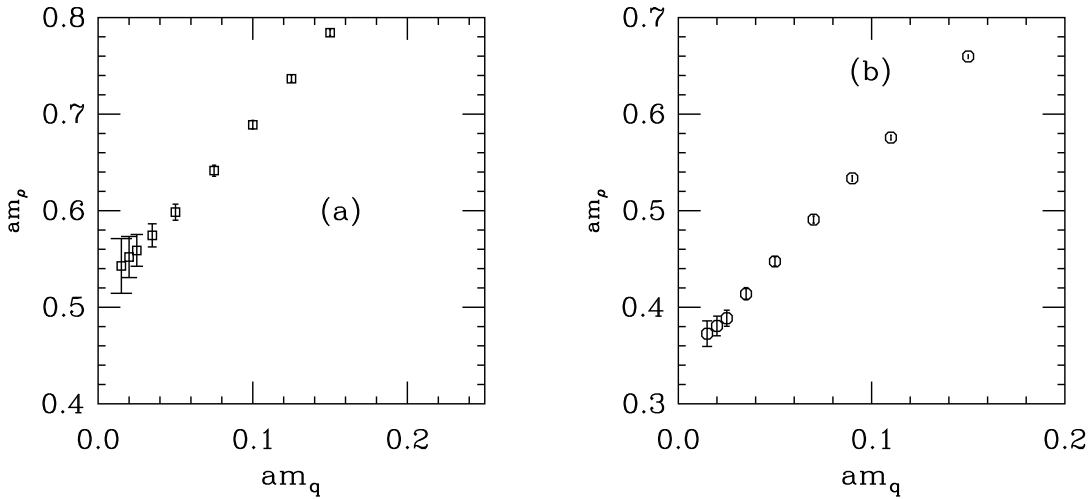}
\end{center}
\caption{$am_{V}$ vs $am_q$, (a) $\beta=5.9$, (b) $\beta=6.1$.
}
\label{fig:rho85.9}
\end{figure}

The vector meson mass is shown as a function of the quark
 mass (both in lattice units) in Fig. \ref{fig:rho85.9}). It was
measured with the correlator of two vector currents. It is expected
 to have only an ``interference term'' like
the axial vector correlator.
 As the quality of the signal degrades at low quark mass, and as
I only use vector masses at larger quark masses, or use
all masses to make linear extrapolations to the chiral limit, I
will not investigate this data set further.

\begin{figure}[thb]
\begin{center}
\epsfxsize=0.8 \hsize
\epsffile{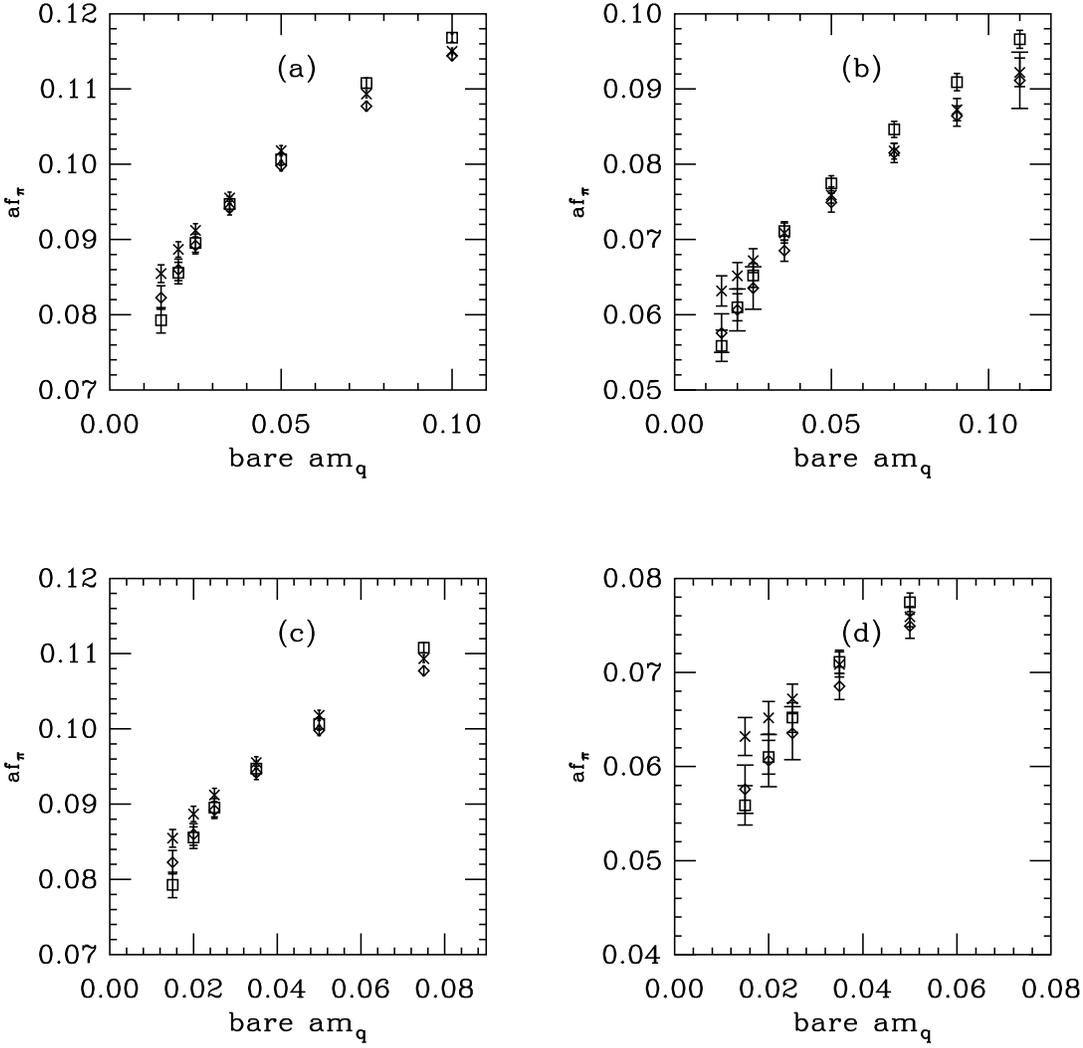}
\end{center}
\caption{ (a)
$(af_{PS})$ vs $m_q$, (a) $\beta=5.9$, (b) $\beta=6.1$.
The labels crosses, squares, and diamonds
 label decay constants extracted from
pseudoscalar correlators, differences of pseudoscalar
 and scalar correlators, and
correlators with
axial sources.
(c), (d): The small mass region of panels (a), (b) magnified.
}
\label{fig:fpib}
\end{figure}

The pseudoscalar decay constant,
defined through 
 $f_{PS}=m_{PS}^2/(2m_q) \langle 0| \bar \psi \gamma_5 \psi | PS \rangle$,
 is shown in Fig. \ref{fig:fpib}.
 It is extracted it from a correlated fit to
a two-point function with Gaussian source and sink and one with a
Gaussian source and local pseudoscalar current sink.
I have used pseudoscalar and axial sources, as well as a fit where the two
 correlators are the pseudoscalar-scalar difference.
There is some tendency for the latter correlators to
 produce a smaller decay constant at smaller
quark mass and a larger one at larger quark mass. The
deviation of the decay constant measured using pseudoscalar sources and sinks
from that measured in other channels at lower quark mass
 may reflect the effects of zero modes in that channel (it is consistent with
 previous work by\cite{Giusti:2001pk}).
The deviation of the quantity measured in the
pseudoscalar-scalar difference channel
at larger quark mass is almost certainly
 due to the degradation in the signal
from the difference correlator there: the fit simultaneously
 produces the pseudoscalar mass along with the
decay constant, and this mass also drifts away from that of
 the other channels.

When  $(af_{PS})$is needed below, I will use the
 pseudoscalar-scalar difference at low quark mass
and the axial source correlators at higher mass, and vary the
 crossover mass to make sure
 that the results are not sensitive to it.

\begin{figure}[thb]
\begin{center}
\epsfxsize=0.8 \hsize
\epsffile{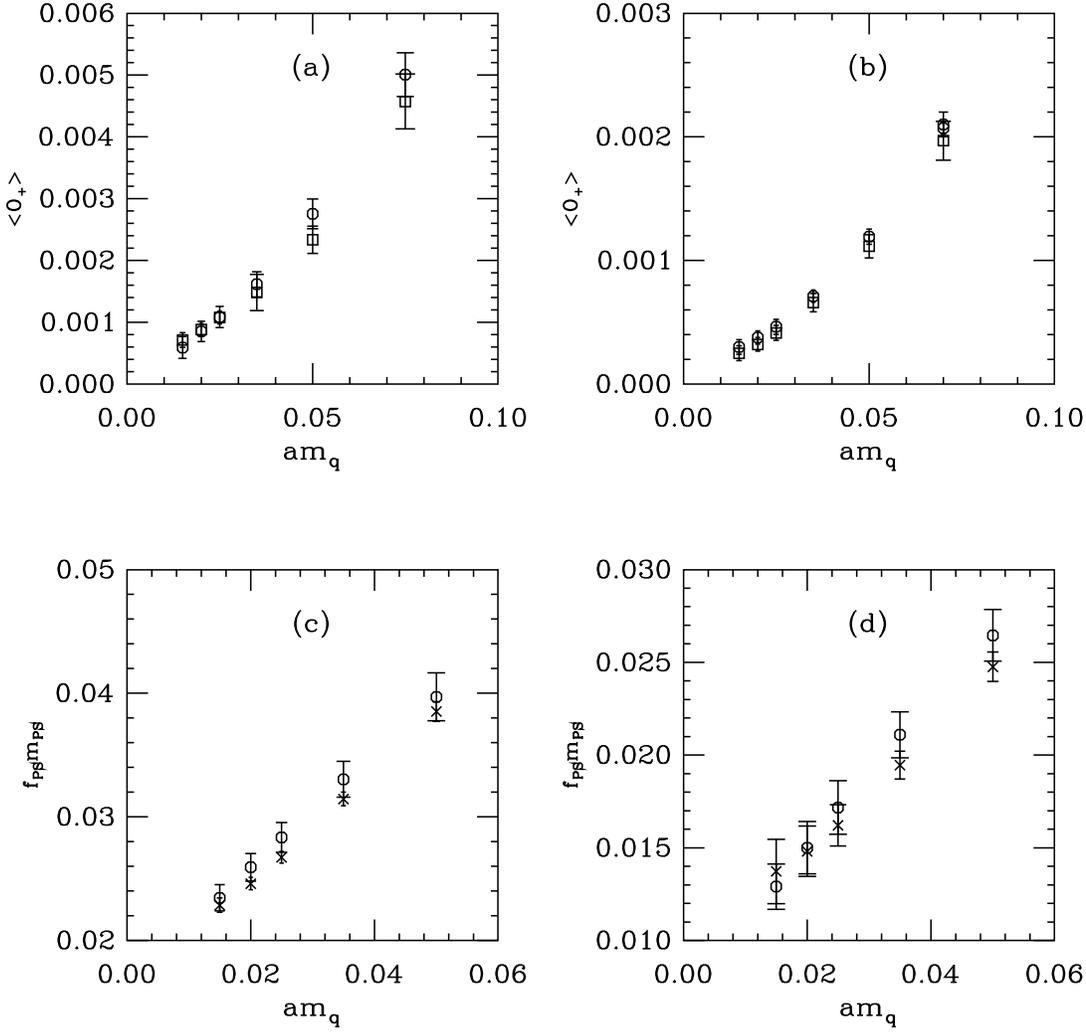}
\end{center}
\caption{Three  parameter correlated fits extracting the matrix element
 $\langle O_+\rangle$ from
a correlator of the operator and a two point function.
 (a) $\beta=5.9$, (b), $\beta=6.1$.
Meson interpolating fields are squares for pseudoscalar-scalar
 difference source
and octagons for the axial vector source for both correlation functions.
The ``denominator'' of $B_K$ can be checked by extracting the
axial vector current matrix element
$m_{PS}f^A_{PS}$ from
a correlated fit of two point functions; this is shown in  (c) and (d) for
$\beta=5.9$ and 6.1. Here crosses show the pseudoscalar-axial vector correlator
and octagons from the difference of $(\gamma_0\gamma_5-\gamma_0\gamma_5)$  and
$(\gamma_0-\gamma_0)$  correlators, which has no zero modes.
}
\label{fig:brat610}
\end{figure}

Finally, there are the operators contributing to $B_K$ itself.
I performed correlated  fits of three and two point
functions (Eqs. \ref{eq:p32a} and \ref{eq:p32b}) to extract the
 matrix element of $O_+$. 
I considered two cases: pseudoscalar sources for the three
 point function, and the pseudoscalar - scalar difference for the two point
 function, and fits to correlators where all interpolating fields are axial currents.
Differences in the fits under variation in the ranges
 are comparable to the error bars shown.
I also investigated fits to the axial vector current operator, using
source interpolating fields which were either pseudoscalar or axial vector.
An example of such a search is shown in Fig. (\ref{fig:brat610}).
Only at the smallest mass, not at the  kaon, does the
 operator appear to be affected.
Thus the extraction of the $B$ parameter in the
 chiral limit might be compromised by finite volume effects, but not
$B_K$ itself.

%

\section{Data analysis for $B_K$}

\subsection{Fits to Lattice $B_K$}

\begin{figure}[thb]
\begin{center}
\epsfxsize=0.8 \hsize
\epsffile{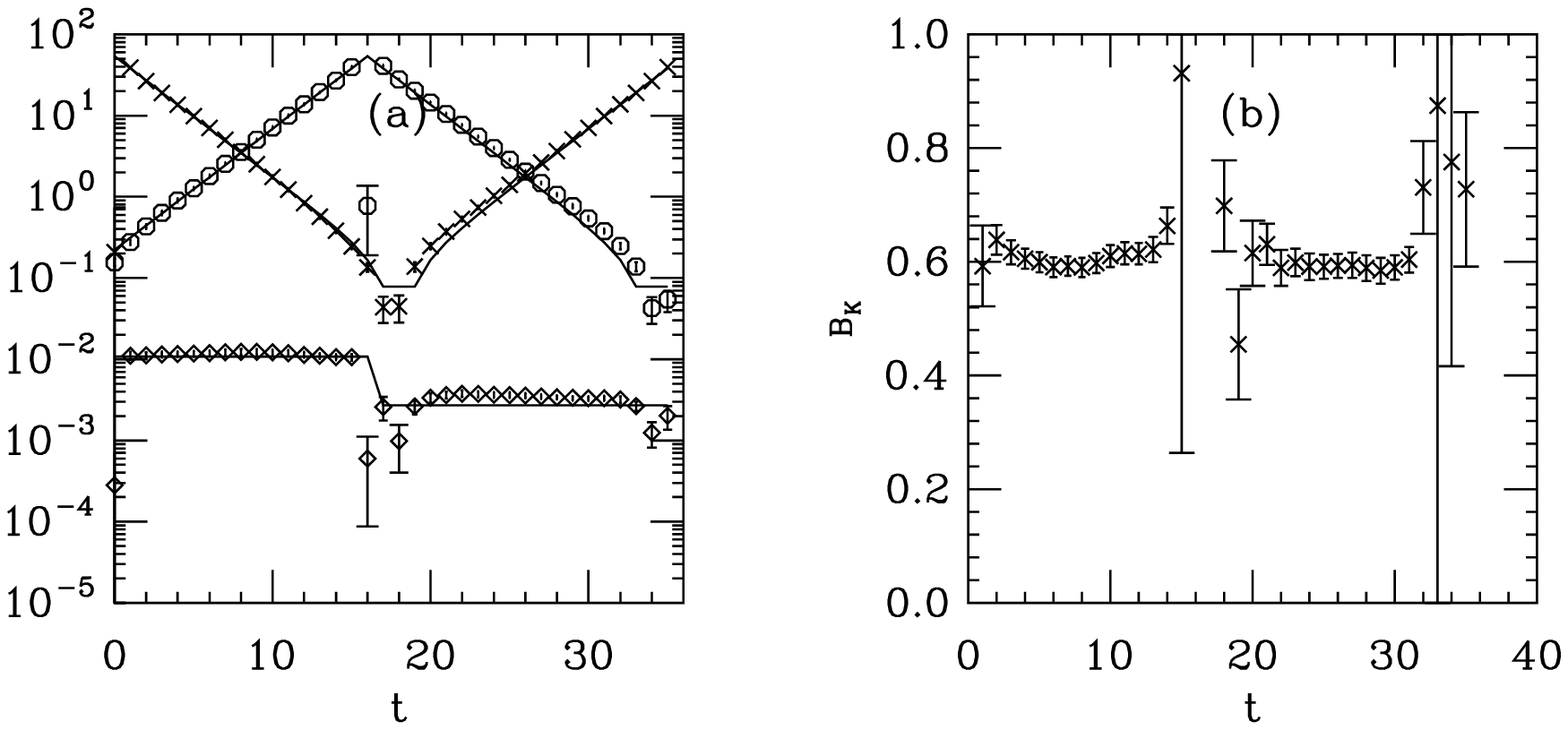}
\end{center}
\caption{Examples of data sets, $\beta=5.9$, $am_q=0.035$, PS sources.
(a) Correlators from which 
$B_K$ is fit, with solid lines showing the fit.
Correlators 1 and 2 (crosses and octagons) have pseudoscalar sources and
axial vector sinks; correlator 3 (diamonds) is the
figure-eight graph.
(b) Ratio of correlators from which
$B_K$ is fit.
}
\label{fig:example}
\end{figure}

\begin{figure}[thb]
\begin{center}
\epsfxsize=0.8 \hsize
\epsffile{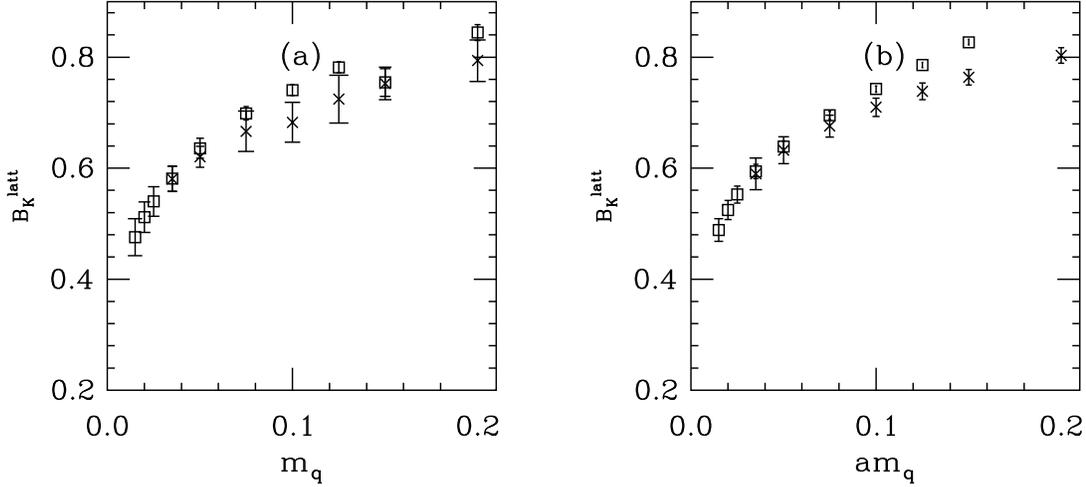}
\end{center}
\caption{$B_K$ extracted from (a) correlated fits to three propagators
and (b) jackknifed ratios of propagators, with  pseudoscalar
 sources, $\beta=5.9$. Fits are over the range $t=6-10$, 24-28. In (a),
squares show three-parameter fits and crosses show four-parameter fits,
while in (b) the squares are from fits of the ratio to a constant, while the
crosses include the lowest $\vec p \ne 0$ contribution.
}
\label{fig:bkjack5.9}
\end{figure}

\begin{figure}[thb]
\begin{center}
\epsfxsize=0.8 \hsize
\epsffile{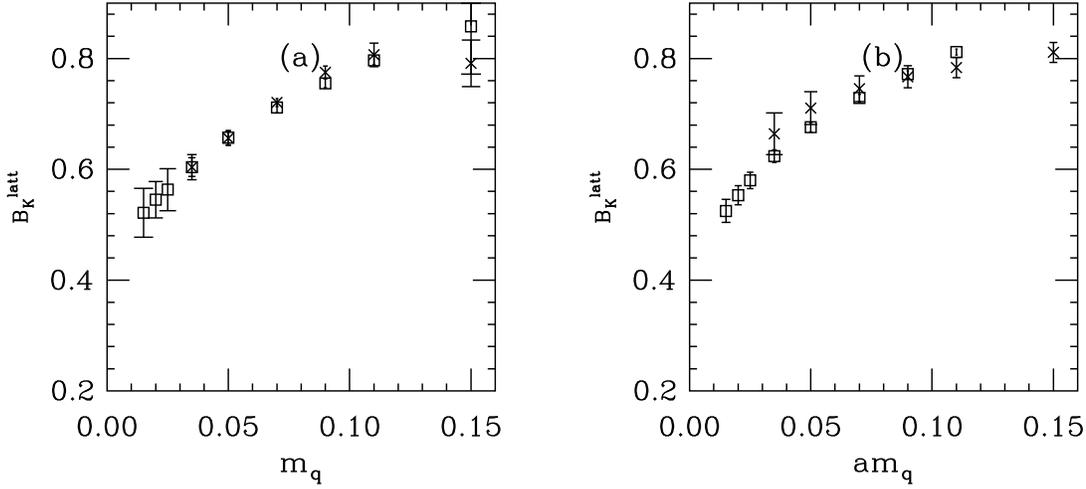}
\end{center}
\caption{$B_K$ extracted from (a) correlated fits to three propagators
and (b) jackknifed ratios of propagators, with  pseudoscalar 
 sources, $\beta=6.1$.  sources, $\beta=5.9$.
 Fits are over the range $t=8-14$, 30-38. Labels are as in
 Fig. \protect{\ref{fig:bkjack5.9}}.
}
\label{fig:bkjack6.1}
\end{figure}

\begin{figure}[thb]
\begin{center}
\epsfxsize=0.8 \hsize
\epsffile{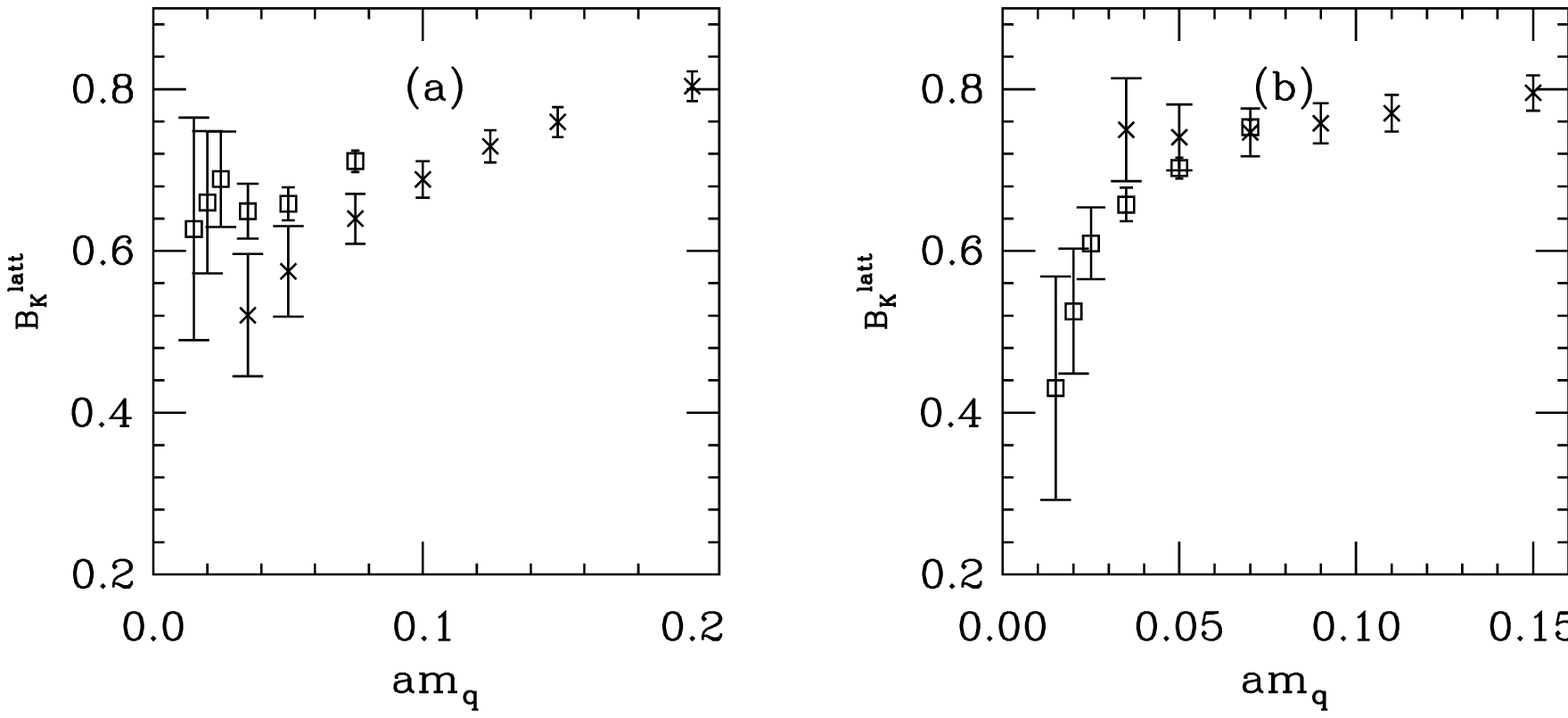}
\end{center}
\caption{$B_K$ extracted from jackknife averages of ratios of
correlators, with  axial vector sources,  (a)  $\beta=5.9$, (b) $\beta=6.1$.
Squares show constant-ratio fits and crosses show  fits keeping the $\vec p\ne 0$ term.
}
\label{fig:bkjacka}
\end{figure}

$B_K$ can be extracted  from covariant fits or ratios of three propagators.
A typical  correlated fit of the figure-eight graph and
two axial vector sink correlators,
and a ratio plot are is shown in Fig. \ref{fig:example}.
Correlated fits are stable against variations in fit ranges or cuts on the
correlation matrix and  have reasonable confidence levels.
Uncorrelated jackknife ``ratio fits'' are consistent with the correlated fits, have small
uncertainties and are also quite stable
over a wide range of time slices.
Examples of these extractions are shown in
 Figs.  \ref{fig:bkjack5.9},  \ref{fig:bkjack6.1} and \ref{fig:bkjacka}.
My results for the B-parameter in lattice regularization as
a function of the bare quark mass are shown in Tables
\ref{tab:bk5.9} and \ref{tab:bk6.1}.

%
%

\subsection{Matching Factors}
 The naive dimensional regularization (NDR) $B_K$ at
 a scale $\mu$ is related to the lattice-regulated
number (from lattice scale $a$) by
\bee
B_K^{(NDR)}(\mu) = R B_K^{(latt)} = {{Z_{TOT}(\mu,a)}\over Z_A^2} B_K^{(latt)}
\ee
where the conversion factor for the operator $O_+$, $Z_{TOT}$, can
 be decomposed into
\bee
Z_{TOT}(\mu,a) = Z_{NDR}(\mu,m) Z_{PT}(ma,qa).
\ee
$Z_A$ is the axial current matching factor.
$Z_{NDR}(\mu,a)$ is the usual two-loop running formula for
 the NDR operator, carrying it from scale $m$
to scale $\mu$. $Z_{PT}(ma,qa)$ is the lattice-to-NDR
 matching factor connecting lattice scale $a$
with continuum scale $m$,
\bee
Z_{PT}(ma,qa)= 1 + {{g^2(qa)}\over{16\pi^2}}[-\gamma_0 \ln{ma} + b]
\ee
$\gamma_0$ is the anomalous dimension of the operator;
 $b$ is the lattice-action-dependent matching term
($b= -3.98$ for the action used here \cite{DeGrand:2002va}).
 One can imagine several different choices
for the product of $Z$'s.

\begin{itemize}
\item{$ma=1$, $q=q^*$ as defined by the Lepage-Mackenzie
 matching condition \cite{ref:LM}.
I will call this``type 1.''}
\item{Gupta, Bhattacharya, and Sharpe\cite{Gupta:1996yt} suggest
 two versions of ``horizontal matching,''
where $m=q_i$, $q^*=q_i$. and $q_i a=1$ (type 2) or $\pi$ (type 3).}
\end{itemize}
The RGI (renormalization group invariant) $\hat B_K$ is
 computed from the NDR B-parameter
in the usual way,
and I will write the constant which renormalizes the
 lattice number as $\hat B_K = \hat R B_K^{(latt)}$.

I determine the strong coupling constant (in the so-called
 $\alpha_V$ prescription)
 at a scale $q=3.41/a$ from the plaquette, convert it
to $\overline{MS}$ prescription, and run it to the
desired scale using the two-loop evolution equation. 
(In all cases I determine $Z_A$ perturbatively, 0.988 at
 $\beta=5.9$, 0.990 at 6.1.)
At $\beta=5.9$ the ``type 1'' Z-factor is $R=0.97$ and the
 type 2 choices are 1.00 and 0.95.
The conversion factors to the RGI $\hat B$ are 1.35, 1.40,
 1.31 for the three choices.
The $\beta=6.1$ numbers are almost equal:
 for $B_K^{(NDR)}(\mu=2$ GeV), 0.98, 1.01,0.97 
and for $\hat B$, 1.38, 1.42, 1.36.
It appears that a reasonable choice and uncertainty
 for $R$ is 0.97(3) at $\beta=5.9$ and
0.98(3) at 6.1.
(Note that in an extrapolation to the continuum, the same choice for
matching convention must be used at all lattice spacings.)

\subsection{``Quenched phenomenology'' and the strange quark mass}
$B_K$ comes from a meson made of two degenerate
quarks with a common mass which
is the average of the nonstrange and strange quark mass.
Unfortunately, if quenched spectroscopy is
 different from the spectroscopy of full QCD, there is no unique
choice for this quark mass,
 even in the absence of discretizaton-induced scale violations.
What follows is a discussion which lies outside the realm
of lattice calculations, as I try to motivate some choices
for lattice spacings and quark masses over others.
Unfortunately, Figs \ref{fig:bkjack5.9}-\ref{fig:bkjack6.1}
show a rather large dependence of the $B$ parameter on the quark
mass at small quark mass, so the choice of mass
 is important phenomenologically.

One choice would be to take the lattice spacing from the
 Sommer parameter (using the interpolation
formula of Ref. \cite{ref:precis}) and
 to set the pseudoscalar mass at the kaon.
Either interpolation or a quadratic fit to the pseudoscalar mass
 $m_{PS}^2 =d m_q$  to $m_K r_0=1.256$  gives $am_s/2=0.027(1)$
 at $\beta=5.9$, 0.019(1) at 6.1.  (The reader might recall 
that the interpolation formula of Ref. \cite{ref:precis} plus
a nominal value of $r_0=0.5$ fm implies that $a$ would be
 0.11 or 0.08 fm at $\beta=5.9$ or 6.1.)

 However, setting the lattice spacing from $r_0$ will push 
 the predictions for other physical quantities away from
experiment. For example, one can extrapolate the vector meson
mass, the pseudoscalar
decay constant, and their ratio, into zero quark mass. 
(Both extrapolations are linear in $m_q$; in contrast to full QCD, the quenched decay
constant has a vanishing chiral logarithm.)
 A jackknife extrapolation
 of $f_{PS}/m_\rho$ 
gives 0.166(6) at $\beta=5.9$ and 0.169(8) at $\beta=6.1$, quite close
to the physical value of (132 MeV)/(770 MeV)=0.17.
Separate extrapolations of $m_V$ and $f_{PS}$ to zero quark mass
 would combine with the Sommer parameter to produce too small values
of the rho mass or pseudoscalar decay constants:
jackknife extrapolations of the former quantity give
$am_\rho=0.51(2)$ at $\beta=5.9$ and 0.35(1)) at $\beta=6.1$,
while $af_{PS}= 0.084(2)$ or 0.060(2). 
This means that a better match to non-$B_K$ phenomenology requires
using either
 $f_\pi$ or $m_\rho$ to set the scale.

I tried a variety of fits and dimensionless ratios to find a quark
mass for the kaon.
Two choices which produce similar (but not identical) values are
(1) Take the lattice spacing from the rho mass; interpolate
or fit $m_{PS}^2$ vs $m_q$ over a range of quark masses around the kaon
 to the kaon mass.
(2) Simultaneously determine $a$ and $m_s$ from the phi meson
(assume linear variation of the vector meson mass on the quark mass
and the ``strange eta''
  $\bar s s$ meson: $m_{\eta_s}^2 + m_\pi^2=
2m_K^2$); then shift the quark mass:
 ($m_s + m_{ns})/m_s = 2m_K^2/m_{\eta_s}^2$.
Both fits are taken over a range
 of quark masses between $1/2 m_s$ and $3/2 m_s$.
One can substitute polynomial interpolation for fitting with no
change in the resulting quark mass. The first choice gives $am_q=0.035(2)$
at $\beta=5.9$, 0.025(2) at $\beta=6.1$; the second choice,
 $am_q=0.029(2)$
at $\beta=5.9$, 0.023(2) at $\beta=6.1$.
The values of the $\mu=2$ GeV $\overline{MS}$ strange quark masses from these two choices
110(10) or 92(10)  MeV at $\beta=5.9$
and 112(7) or 104(8) MeV at $\beta=6.1$.
 There is no way to
 determine if the difference in these numbers
is statistical or a quenching systematic. 

\begin{figure}[thb]
\begin{center}
\epsfxsize=0.8 \hsize
\epsffile{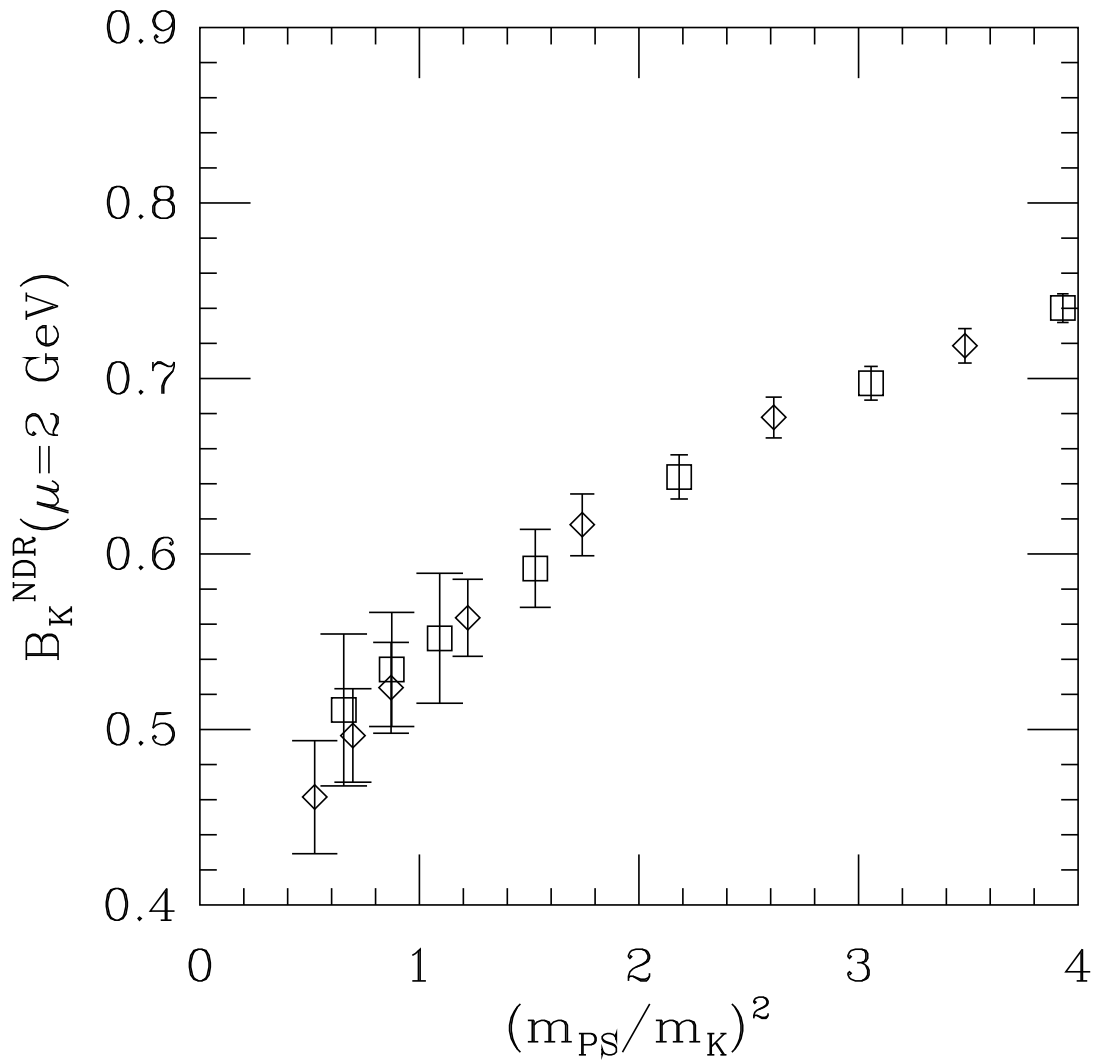}
\end{center}
\caption{ $B_K^{NDR}(\mu=2$ GeV)
 as a function of quark mass from the two data sets--diamonds for $\beta=5.9$,
squares for $\beta=6.1$. The matching factor 
uses the Lepage-Mackenzie convention and the quark mass comes from a combined fit to
 the $\phi$ and $\bar s s$
pseudoscalar.}
\label{fig:combinedbk}
\end{figure}

A comparison of $B_K^{NDR}(\mu=2$ GeV)
as a function of the pseudoscalar mass in units of the kaon mass for
this choice of quark mass ($(m_{PS}/m_K)^2=m_q/((m_s+m_d)/2)$) is shown in Fig.
\ref{fig:combinedbk}. The two data sets appear to scale reasonably well.

In practice, I perform the interpolation of the B-parameter
to the kaon under the same jackknife as the determination of the
strange quark mass. Both determinations have a statistical uncertainty,
and the jackknife attempts to include their correlations. I tried
two methods of interpolation: I fit the data of the lowest
five quark masses to the chiral logarithm form (Eq. \ref{eq:bkm}, below)
or just did a simple polynomial interpolation. Both results agree within
uncertainties, although at $\beta=6.1$ the results
of the chiral fit produce unacceptably large
uncertainties because the chiral parameters are poorly determined.
I will use the interpolated values.

To be definite,  make choice (2) for the quark mass and take the
Lepage-Mackenzie choice for the $Z$-factor.
In that case, one will have $B_K^{NDR}(\mu=2$ GeV)$=0.540(25)$ at $\beta=5.9$,
0.546(36) at $\beta=6.1$.
They lie a bit lower than the staggered
JLQCD result\cite{Aoki:1997nr} or the other overlap
calculation of Garron, et al\cite{Garron:2003cb},
but bracket the two domain wall results of
Refs. \cite{AliKhan:2001wr} and \cite{Blum:2001xb}.
The alternative ``choice (1) quark mass'' would give $B_K=0.563(23)$,
0.552(28)  for the two couplings,

Recall that Ref. \cite{Blum:2001xb}
sets $a$ from the rho mass and chooses the bare quark
 mass to put $m_K$ at ``its physical mass''
while Ref. \cite{AliKhan:2001wr}
sets the strange quark mass
``from the experimental value of $m_K/m_\rho$.''
My neglect of $r_0$ as compared to hadronic masses is at least consistent
with the choice of groups with which I want to compare. (Of course, that
does not make it correct.)
Oddly enough, the ``choice 1 mass'' which gives us a larger
$B_K$ is more like the RBC convention (and their  value for $B_K$ is below mine),
and the ``choice 2 quark mass''  which gives a lower mass is more like that of
Ref. \cite{AliKhan:2001wr}, which lies above my data.
Perhaps I am just being obsessive about one standard deviation effects.

Although they show no scaling violations, an extrapolation of the
``type 2 mass'' results  to
zero lattice spacing $B(a)= B_0 + a^2 B_1$ would give a continuum
prediction of $B_K=0.550(67)$ for this choice of quark mass and
 matching factor. (The RGI $\hat B_K$ would be
0.751(35) at 5.9,
0.769(51) at 6.1
and 0.781(94)
for an extrapolated continuum value.)
These values are the points shown in
 Figs. \ref{fig:worldbka} and \ref{fig:worldbk}.
The alternative ``choice (1) quark mass'' would give $B_K= 0.544(53)$
 for the continuum value,
and any other of the choices for the perturbative $Z$ factor would produce
a continuum result whose central value is also within 0.01 of $B_K=0.55$
with an  extrapolation uncertainty of 0.05 ($\hat B_K=0.78(8))$.
The  variation in $B_K$ at fixed lattice spacing from varying the choice of $Z$ is completely
washed out by the extrapolation.

These results are quite similar to continuum $1/N_c$ predictions of $\hat B=0.7(1)$
\cite{Bardeen:1987vg}.

\section{Results in the chiral limit: $O_+$, $O_7$ and $O_8$}
\subsection{$O_+$}
\begin{figure}[thb]
\begin{center}
\epsfxsize=0.6 \hsize
\epsffile{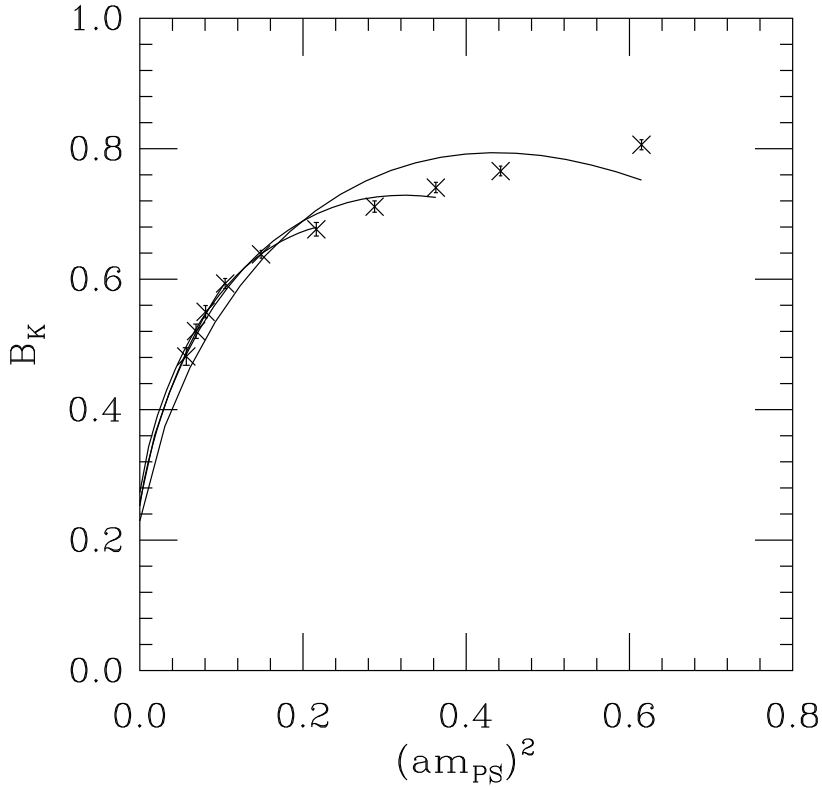}
\end{center}
\caption{
A set of chiral extrapolations over various mass ranges
 (shown by the extent of the solid lines)
to lattice $B_K$, from jackknifed ratios of correlators at $\beta=5.9$ .
}
\label{fig:typical2pch}
\end{figure}

The behavior of $B_K$ as a function of pseudoscalar meson mass
 has been computed by
Sharpe\cite{Sharpe:1992ft}.
 His result (simplified to the degenerate-mass limit
 appropriate to these simulations) is
\bee
B(m)= B[ 1 - {{3 m_{PS}^2} \over {8\pi^2 f_\pi^2}}
\log{m_{PS}^2} ] + c m_{PS}^2
\label{eq:bkm}
\ee
I follow the path of Ref. \cite{Blum:2001xb} and
 perform a chiral extrapolation using
the functional form of Eq. \ref{eq:bkm}. 
The range of validity in quark mass of this formula
 is not known a priori, and so I
simply performed fits of my data to it dropping one
 heavy quark mass at a time, and looking for stability in the extrapolated
result.
The fit involves a jackknife to incorporate the
 measured pseudoscalar mass, B-parameter, and
$f_\pi$.
 A ``typical''
set of fits (one of the ones in a jackknife ensemble)
is shown in Fig. \ref{fig:typical2pch}.
I find that the quality of the fit is good,
and extrapolated value of $B$ is insensitive to the
 number of masses kept, as long
as I keep less than about 8 of the masses (this
 corresponds to a pseudoscalar/vector mass ratio $\le 0.75$)
I find  extrapolated  ``type 1'' values of $B=0.27(2)$ at $\beta=5.9$;
0.23(4) at $\beta=6.1$ (or $\hat B=0.36(2)$, 0.31(7)).
This is in reasonably good agreement with the more precise RBC number of $B=0.267(14)$.

The authors of Ref. \cite{AliKhan:2001wr} performed a
 chiral extrapolation of their data
letting the coefficient of the chiral logarithm be a
 free parameter. Their result (0.420(16))
is much larger than mine. I believe that this is the
 result of using a different functional
form to extrapolate. I believe that using the known
 analytic functional formula for chiral 
extrapolation is theoretically better justified.

\subsection{$O_7$ and $O_8$}

My conventions for the operators and matrix elements are the same as
those of Refs. \cite{Blum:2001xb} and \cite{Noaki:2001un},
 and have been given in Eqs.
\ref{eq:o7} and \ref{eq:o8}. In the degenerate quark mass
 limit where I work, matrix elements
of these operators only involve figure-eight graphs. The
 physical parameters one wishes
 to associate with these operators are the zero-quark mass
 limit of the matrix elements,
and so a chiral extrapolation is needed. $O_7$ and $O_8$
 also mix with each other under renormalization.
The mixing matrix was computed in one loop perturbation
 theory in Ref. \cite{DeGrand:2002va}.
I compute the NDR matrix element by extrapolating each
 lattice operator to the chiral limit,
then mixing and running the operators, all inside a single
 elimination jackknife.

The matrix element of these operators has a chiral
 extrapolation similar to that of $B_K$,
\bee
< \pi |O_i | K> = C (1 + {{\xi m_{PS}^2 } \over {(4\pi f)^2}} \ln (m_{PS}^2) )
   + b m_{PS}^2 .
\ee
At the time that Ref. \cite{Blum:2001xb} appeared, the
 size of the chiral logarithm for these
operators was not known, and the authors did fits which
 either dropped the logarithm
or left it as a free parameter. Since then, Golterman
 and Pallente \cite{Golterman:2001qj}
have computed the coefficients for chiral logarithms 
in quenched and partially-quenched QCD,
and found that $\xi=0$ for $O_7$ and $O_8$ (in the
 degenerate-mass limit  used here).
Accordingly, I just extrapolate the data to zero
 assuming a linear dependence on the quark mass.

A peculiarity of the quenched approximation is that
 the B-parameters for these operators
go to zero in the chiral limit, because $m_{PS}^2/m_q$
 diverges. Recall that the B-parameters
for these operators are defined as
\bee
B_i^{3/2} = {{\langle K|O_i^{3/2}|\pi\rangle} \over{
c_P \langle \pi | \bar \psi \gamma_5 \psi |0 \rangle
 \langle 0 | \bar \psi \gamma_5 \psi |K \rangle
+ c_A \langle \pi | \bar \psi\gamma_\mu \gamma_5 \psi |0 \rangle
 \langle 0 | \bar \psi \gamma_\mu\gamma_5 \psi |K \rangle }}
\ee
where the $c_i$'s are numerical coefficients.
The PCAC relation says that
 $\langle 0 | \bar \psi \gamma_5 \psi |PS \rangle ={1\over 2}
  m_{PS}^2/m_q f_{PS}$.
The divergence of $(m_{PS}^2/m_q)^2$ in the denominator is
 not compensated by any singular behavior in the
numerator.

Therefore, I focus on the matrix
 elements themselves. Several forms have been presented
in the literature; $\langle K | O_i |\pi\rangle$ itself,
\bee
\langle K | O_i |\pi\rangle = 12{\alpha_i \over f_\pi^2}
\label{eq:alph}
\ee
the lowest order (in chiral perturbation theory)
 $K\rightarrow (\pi\pi)_{I=2}$ matrix
element
\bee
M_i(K\rightarrow (\pi\pi)_{I=2}) = {{\langle K | O_i |\pi\rangle}\over f_\pi}
\label{eq:mmmm}
\ee
and a  dimensionless coupling constant, defined as
\bee
g((K\rightarrow (\pi\pi)_{I=2})={{\langle K | O_i |\pi\rangle}\over f_\pi^4}.
\label{eq:gggg}
\ee
Note $O_i = O_i^{latt} (1/a)^4$, $\alpha_i = \alpha_i^{latt} (1/a)^6$, and $M_i = M_i^{latt} (1/a)^3$.

\begin{figure}[thb]
\begin{center}
\epsfxsize=0.8 \hsize
\epsffile{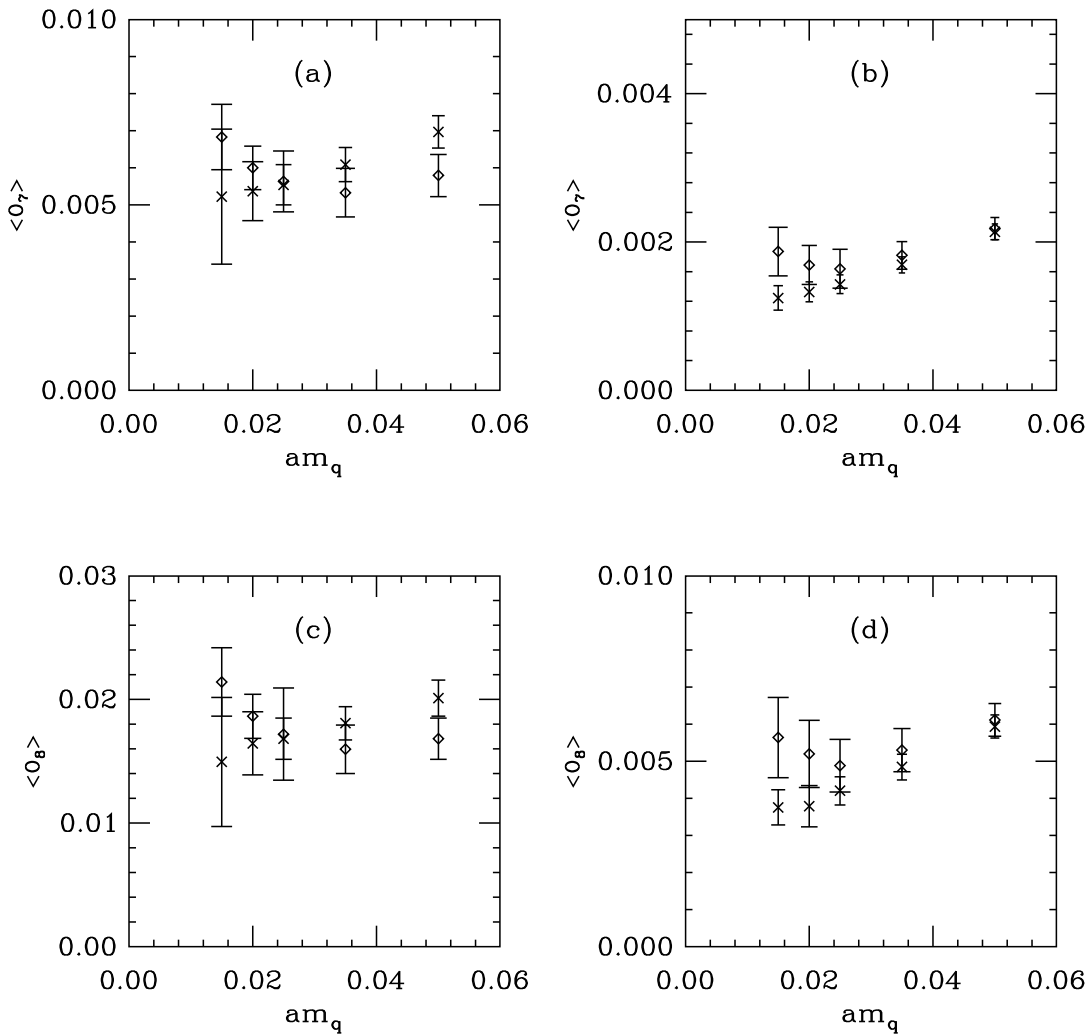}
\end{center}
\caption{ (a) ($\beta=5.9$) and (b) ($\beta=6.1$)
 $\langle O_7 \rangle$ and
(c) ($\beta=5.9$) and (d) ($\beta=6.1$)
 $\langle O_8 \rangle$ 
from 
(diamonds)  pseudoscalar sources and sinks in the
three point function and the pseudoscalar-scalar difference
 for the two point function, and (cross) axial
source and sink for both correlators.
}
\label{fig:o78610}
\end{figure}

Because one is interested in the matrix elements themselves,
 I extract them from correlated
fits to a three-point (figure-eight) function and a
 two point function with the same source and sink. As before,
I have measured three point correlators with pseudoscalar
 and axial vector Gaussian sources.
I expect these operators might show different zero mode
 effects. In Fig. \ref{fig:o78610}
I show $\langle O_7 \rangle$ and  $\langle O_8 \rangle$
 for two choices of source/sink
(pseudoscalar and axial) for the three point function,
 and pseudoscalar sources and sinks, 
the pseudoscalar-scalar difference
 for the two point function, and  axial
source and sink for both correlator. Only at the lightest
 quark masses are there  noticible differences.
 I have performed jackknife extrapolations of the matrix elements
to the chiral limit, using the lightest five quark masses and
taking a simple linear mass dependence.
 When I do that, I find
that  the uncertainty on the
extrapolated matrix elements inflates so much, that  no dependence on
 the choice of sources and sinks is seen.

\begin{figure}[thb]
\begin{center}
\epsfxsize=0.8 \hsize
\epsffile{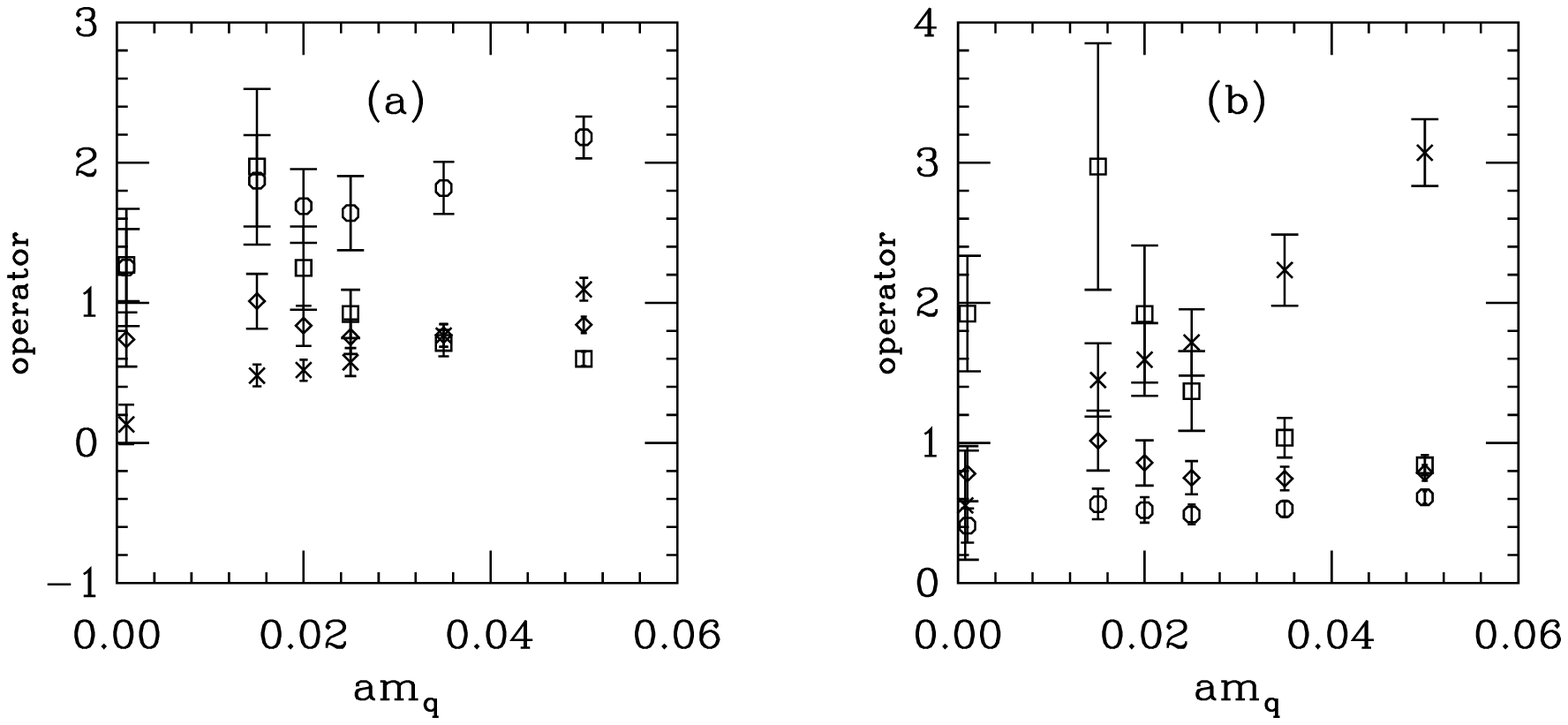}
\end{center}
\caption{
Various scaled versions of $O_7$ and $O_8$ from the $\beta=6.1$ data set.
The data shown at the origin are the extrapolated values
(a)
$\langle K | O_7 |\pi\rangle$, scaled by   3000  (octagons);
$\alpha_7$, from Eq. \protect{\ref{eq:alph}}, scaled by $10^6$(crosses);
$M_7$, from Eq. \protect{\ref{eq:mmmm}}, scaled by 30 (diamonds);
$g_7$, from Eq. \protect{\ref{eq:gggg}}, scaled by 0.01 (squares).
(b)
$\langle K | O_8 |\pi\rangle$, scaled by   100  (octagons);
$\alpha_8$, from Eq. \protect{\ref{eq:alph}}, scaled by $10^6$(crosses);
$M_8$, from Eq. \protect{\ref{eq:mmmm}}, scaled by 30 (diamonds);
$g_8$, from Eq. \protect{\ref{eq:gggg}}, scaled by 0.005 (squares).
}
\label{fig:comp78}
\end{figure}
I extrapolate the alternative versions of the matrix elements by including
the calculation of the pseudoscalar decay constant in the jackknife and extrapolation.
(For example, the $M_i$'s are found by extrapolating $\langle O_i (m_q) \rangle /f_{PS}(m_q)$.)
The matrix elements and decay constants are both gently falling functions of the quark mass, and
their values are strongly correlated in the data set. Thus, different combinations will extrapolate
differently. My experience is illustrated with the noisier $\beta=6.1$ data set (Fig. \ref{fig:comp78}):
The $\alpha_i$'s vary strongly with quark mass and the error on the extrapolated value is large.
Ratios of $\langle O_i (m_q) \rangle$ to powers of $f_{PS}$ fare better. However, the high
dimensionality of the operator, $M_i$, and $\alpha_i$ mean that my results for these operators
are particularly sensitive to the choice of lattice spacing. Thus I will restrict myself
to a presentation of the matrix elements themselves and the dimensionless  $g_i$'s,
shown in tables \ref{tab:me5.9}-\ref{tab:me6.1}.

The three choices of matching factors correspond to the following mixing matrices at $\beta=5.9$, 6.1:
Type (1):
\bee
Z = 
\left(\matrix{ 0.979 & 0.005 \cr
                                             -0.027 & 1.099\cr}\right);
\left(\matrix{ 0.979 & -0.060 \cr
                                             -0.019 & 1.018\cr}\right)
\ee
Type (2):
\bee
Z = 
\left(\matrix{ 0.979 & 0.0016 \cr
                                             -0.038 & 1.110\cr}\right);
\left(\matrix{ 0.996 & -0.040 \cr
                                             -0.041 & 1.000\cr}\right)
\ee
Type (3):
\bee
Z = 
\left(\matrix{ 0.996 & -0.026 \cr
                                             -0.042 & 1.021\cr}\right);
\left(\matrix{ 1.010 & -0.059 \cr
                                             -0.044 & 0.934\cr}\right)
\ee
Since the type 1 and 2 Z-factors are so similar, there is a negligible difference in the
results of the  matrix elements from using either. Using the Type 3 Z's suppresses the operators
by ten per cent at $\beta=5.9$, five per cent at $\beta=6.1$.

\section{Conclusions}
Using a lattice action with exact chiral symmetry to do a lattice
calculation
of  a matrix element in quenched
approximation is a two edged sword.
On the one hand, exact chiral symmetry eliminates ``non-continuum''
operator mixing and makes the calculation reasonably straightforward.
On the other hand, because one can do simulations at small values of
the valence quark mass, the artifacts of the quenched approximation
are clearly revealed. It is clear that quenched QCD is not even qualitatively
the same theory as full QCD, and that using the quenched approximation
to compute hadronic matrix elements is merely another kind of phenomenology.

My results show no scale violations. Their continuum extrapolation is the
 prediction for quenched  $B_K^{(NDR)}(\mu=2$ GeV) $=0.55(7)$
or $\hat B=0.79(9)$. This result brackets the
two domain wall predictions of Refs.
 \cite{AliKhan:2001wr}, \cite{Blum:2001xb},
One could improve the statistical uncertainty of this number by
 further simulations,
but I believe that the greatest source of theoretical
 uncertainty is an uncontrolled
systematic error arising from the use of the quenched approximation itself.
I know no way to reliably quantify this systematic.
It would be a far better use of computer resources to devote them to
simulations done in full QCD.

\section*{Acknowledgments}

This work was supported by the US Department of Energy.
I am grateful to S. Sharpe for suggesting this project, to
C. Bernard,
T. Blum,
and to
G. Colangelo
 for helpful instruction,
to M. Golterman for correspondence, and
to the other members of the MILC collaboration for reading and commenting
on the manuscript. Part of this project was
done while I was a guest at the Max Planck Institute for Physics and 
Astrophysics, Munich, and I gratefully acknowledge its hospitality.
 Simulations were performed on the Platinum IA-32 cluster at NCSA.

\appendix
\section{Checking perturbation theory}
It is reasonably straightforward to make a nonperturbative
determination of the axial current matching factor $Z_A$
and the lattice-to- $\overline{MS}$ matching factor for the quark mass, $Z_M$,
and check the perturbative calculation of Ref. \cite{DeGrand:2002va}.

\begin{figure}[thb]
\begin{center}
\epsfxsize=0.8 \hsize
\epsffile{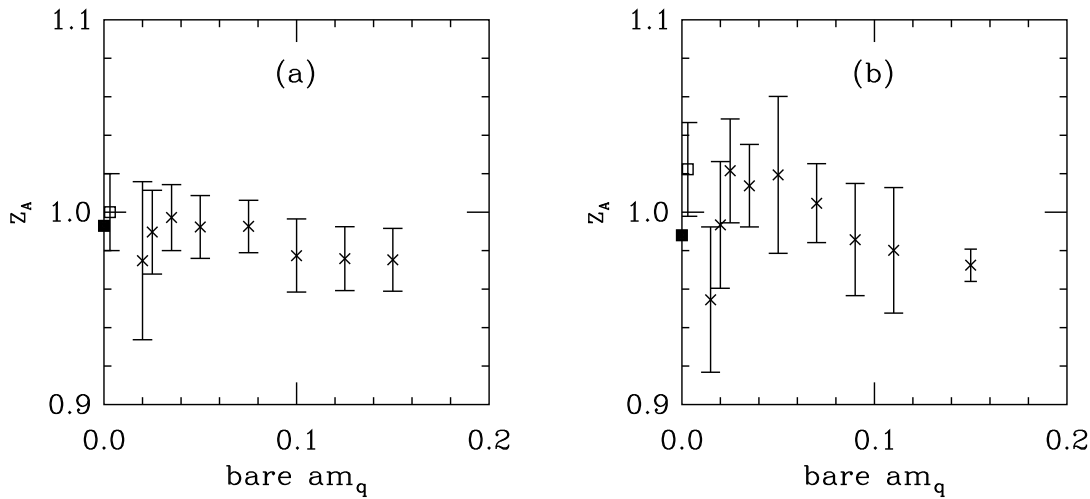}
\end{center}
\caption{Inferred $Z_A$ from ratio of axial vector to pseudoscalar
matrix elements, (a) $\beta=5.9$ (b) $\beta=6.1$.
 The black square is the perturbative prediction.
The open square is the jackknife extrapolation to the massless limit.
}
\label{fig:za5.9}
\end{figure}

I determine $Z_A$ as follows: The matrix element of the pseudoscalar
current gives 
$\langle 0 | \bar \psi \gamma_5 \psi |PS \rangle= f_{PS}^Pm_{PS}^2/(2m_q)$,
with no lattice-to-continuum renormalization factor for $f_{PS}^P$;
$f_{PS}^P=f_{PS}$.
The zeroth component of the axial current has as its matrix element
$\langle 0 | \bar \psi \gamma_0 \gamma_5 \psi |PS \rangle= f_{PS}^Am_{PS}$,
with $f_{PS}^A = Z_A f_{PS}$. Thus $Z_A=f_{PS}^P/f_{PS}^A$.
I compute the ratio by fitting a correlator with a space-summed pseudoscalar
sink and a correlator with an axial vector sink, taking the ratio,
 extrapolating to the chiral limit, and jackknife-averaging.
The result, which is quite insensitive to the range of quark masses
or the choice of timeslices kept in the fit, is 1.00(2) at $\beta=5.9$,
1.02(2) at 6.1.
The perturbative calculation of Ref. \cite{DeGrand:2002va} predicts
 $Z_A=0.973$ or 0.993 at $\beta=5.9$, 0.980 or 0.988 at 6.1
(depending on the choice of lowest-order or higher-order $q^*$; the latter
numbers are ``preferred'' by the ideology of choice of momentum scale
for the strong coupling constant of Ref. \cite{Hornbostel:2000ey}).
$Z_A$ for $\beta=5.9$ and 6.1 is illustrated in Fig. \ref{fig:za5.9}.

The other renormalization constant which has an ``easy'' comparison to
 a nonperturbative calculation
is the matching
coefficient for the $\overline{MS}$ quark mass, as described by 
Ref. \cite{Hernandez:2001yn}.
The method involves determining the value of bare quark mass
at which the pseudoscalar mass takes on a certain value and
combining that bare mass with an appropriately-rescaled renormalization
factor computed using Wilson fermions \cite{Garden:1999fg}, and 
 called $U_m$ by the authors of
Ref. \cite{ref:HJLW}. The $Z-$ factor which converts the lattice
 quark mass
to the RGI (renormalization group invariant) quark mass is 
\bee
\hat Z_M(g_0) = U_m {1\over{r_0 m_q}}|_{(r_0 m_{PS})^2=x_{ref}}.
\label{ZHAT}
\ee

The RGI condensate can then be converted to the $\overline{MS}$ regulated
condensate using a table of (multi-loop) conversion coefficients
from Ref. \cite{Garden:1999fg}:
$Z_{\overline{MS} }(\mu) = \hat Z_M  z(\mu)$
where $z=0.72076$ for $\mu = 2$ GeV.

The authors of Ref. \cite{Garden:1999fg} quote $U$ at $x=r_0^2m_{PS}^2$=5,
 3, 1.5376.
The larger $x$ values occur at quark masses for which $m_{PS}^2/m_q$
 is a gently falling
function (compare Fig. \ref{fig:mpi2mqmq} and Table \ref{tab:npzm}).
 For these values,
polynomial interpolation of the pion mass gives the values shown in the Table.
However, the smallest value of $x$ corresponds to a quark mass deep
 in the region where 
$m_{PS}^2/m_q$ is rapidly rising. To drive the pseudoscalar mass
 to the fiducial value of $x$
requires a quark mass which is ``smaller than it should be''
 (in the absence of a rise),
and so the value of $Z_M$ is enhanced, to about 1.21(8) for
 the $\beta=5.9$ data set.

Note that the authors of Ref. \cite{Garden:1999fg}
 actually did not measure
$U$ at $x=1.5376$; instead, they extrapolated their pseudoscalar
 masses down from higher values
of the quark mass. The extrapolation is done assuming that
 $m_{PS}^2$ is linear in $m_q$,
that is, without chiral enhancements. Simply ignoring
 the lower quark mass data, fit the higher
quark mass data ($am_q=0.035$ to 0.125 at $\beta=5.9$) to
 $a^2 m_{PS}^2 = C am_q$, extrapolate to lower quark mass,
 and using the extrapolated mass in Eq. \ref{ZHAT},
 produces
the results shown in the Table. All three $Z-$ factors seem
 to be consistent, and are slightly
higher than the perturbative prediction of 1.00 at $\beta=5.9$ (the
constant term nearly cancels the logarithm there) and 1.02 at $\beta=6.1$.


\vfill\eject


\begin{table}
\caption{Pseudoscalar mass and lattice-regulated  $B_K$
as a function of bare quark mass, from correlated fits, at $\beta=5.9$.
}
\label{tab:bk5.9}
\vspace{0.05in}
\begin{center}
\begin{tabular}{l|c|c}
\hline
$am_q$ & $am_{PS}$ & lattice $B_K$  \\
\hline
0.015 & 0.227(5)       & 0.476(33) \\
0.020 & 0.254(5)       & 0.512(25) \\
0.025 & 0.278(5)       & 0.540(27) \\
0.035 & 0.320(5)       & 0.581(23) \\
0.050 & 0.386(4)       & 0.636(18) \\
0.075 & 0.467(4)       & 0.667(36) \\
0.100 & 0.537(3)       & 0.683(36) \\
0.125 & 0.603(3)       & 0.724(43) \\
0.150 & 0.665(3)       & 0.753(29) \\
0.200 & 0.784(2)       & 0.794(37) \\
\hline
\end{tabular}
\end{center}
\end{table}

\begin{table}
\caption{Pseudoscalar mass and lattice-regulated  $B_K$
as a function of bare quark mass, from correlated fits, at $\beta=6.1$.}
\label{tab:bk6.1}
\vspace{0.05in}
\begin{center}
\begin{tabular}{l|c|c}
\hline
$am_q$ & $am_{PS}$ & lattice $B_K$  \\
\hline
0.015 & 0.192(5) & 0.512(38) \\
0.020 & 0.218(5) & 0.535(29) \\
0.025 & 0.234(5) & 0.556(47) \\
0.035 & 0.270(4) & 0.599(45) \\
0.050 & 0.319(3) & 0.657(12) \\
0.070 & 0.379(3) & 0.720(10) \\
0.090 & 0.435(2) & 0.771(17) \\
0.110 & 0.488(2) & 0.805(21) \\
0.150 & 0.587(2) & 0.793(50) \\
\hline
\end{tabular}
\end{center}
\end{table}

\begin{table}
\caption{ $\beta=5.9$
atrix elements of operators $O_7$ and $O_8$: first
matrix elements $\langle K| O_i | \pi \rangle$, in units of $a^4$,
$i=7$, 8,
 then 
dimensionless NDR matrix elements $g_i$.
All fits use figure-8 graphs with pseudoscalar sources and ``type 1'' Z-factors.}
\label{tab:me5.9}
\vspace{0.05in}
\begin{center}
\begin{tabular}{l|cc|cc}
\hline
$am_q$ &  $\langle K| O_7 | \pi \rangle \times 10^3$ & $\langle K| O_8 | \pi \rangle\times 10^2$
 &  $g_7$ & $g_8$ \\
\hline
0.015 &   6.83(88)     &   2.14(28)  &    169(38)  &    530(119)         \\
0.020 &   6.00(59)     &   1.86(18)  &    111(17)  &    343(51)         \\
0.025 &   5.63(82)     &   1.72(37)  &     87(14)  &    266(64)         \\
0.035 &   5.33(66)     &   1.60(20)  &     67(10)  &    199(29)         \\
0.050 &   5.79(56)     &   1.68(17)  &     57(6)  &    164(18)         \\
\hline
NDR extrap  &   6.17(90)     &   2.17(33)  &    127(19)  &    433(77)         \\
\hline
\end{tabular}
\end{center}
\end{table}

\begin{table}
\caption{ $\beta=6.1$
atrix elements of operators $O_7$ and $O_8$: first
matrix elements $\langle K| O_i | \pi \rangle$, in units of $a^4$,
$i=7$, 8,
 then 
dimensionless NDR matrix elements $g_i$.
All fits use figure-8 graphs with pseudoscalar sources and ``type 1'' Z-factors.}
\label{tab:me6.1}
\vspace{0.05in}
\begin{center}
\begin{tabular}{l|cc|cc}
\hline
$am_q$ &  $\langle K| O_7 | \pi \rangle \times 10^3$ & $\langle K| O_8 | \pi \rangle\times 10^3$
 &  $g_7$ & $g_8$ \\
\hline
0.015 &   1.87(33)     &   5.64(108)  &    197(56)  &    594(176) \\
0.020 &   1.69(26)     &   5.20(91)  &    125(30)  &    384(98)          \\
0.025 &   1.64(26)     &   4.88(71)  &     92(17)  &    274(57)          \\
0.035 &   1.82(19)     &   5.30(58)  &     71(9)  &    208(28)          \\
0.050 &   2.18(15)     &   6.11(44)  &     60(5)  &    168 15)          \\
\hline
NDR extrap  &   1.08(37)     &   4.04(120)         &    111(22)  &    379(82) \\
\hline
\end{tabular}
\end{center}
\end{table}

\begin{table}
\caption{Nonperturbative determination of $Z_M$}
\label{tab:npzm}
\begin{tabular}{|c|c|c|c|c|}
\hline
$x$ & $U$ & $r_0 m_q$ & $\hat Z_M$ & $Z_{\overline{MS}}(\mu=2 \ \rm{GeV})$ \\
\hline
$\beta=5.9$ &   & & &\\
\hline
5      & 0.580(12) & 0.380(5) & 1.52(4) & 1.10(3) \\
3      & 0.349(9)  & 0.216(5) & 1.62(6) & 1.16(4)  \\
1.5376 & 0.181(6)  & 0.121(1) & 1.50(5) & 1.08(4) \\
\hline
$\beta=6.1$ &   & & &\\
\hline
5      & 0.580(12)       & 0.385(6) & 1.51(4)   & 1.09(3) \\
3      & 0.349(9)       & 0.228(6) & 1.53(5)   & 1.10(4) \\
1.5376      & 0.181(6)       & 0.116(3) & 1.56(6)  & 1.12(4) \\
\hline
\end{tabular}
\end{table}


\end{document}